\documentclass[useAMS,usenatbib]{mn2e}
\usepackage{graphicx}
\usepackage{longtable}
\usepackage{times}

\title[Gaia SPSS survey. I. Preliminary results.]{The Gaia spectrophotometric 
      standard stars survey.\\I. Preliminary results.}
      
\author[E.~Pancino et al.]{E.~Pancino$^{1}$\thanks{email:
   elena.pancino@oabo.inaf.it}, G.~Altavilla$^{1}$, 
   S.~Marinoni$^{1,2,3}$, G.~Cocozza$^{1}$, J.~M.~Carrasco$^{4}$,
   \newauthor
   M.~Bellazzini$^{1}$, A.~Bragaglia$^{1}$, L.~Federici$^{1}$, 
   E.~Rossetti$^{7}$, C.~Cacciari$^{1}$, 
   \newauthor
   L.~Balaguer N\'u\~nez$^{4}$, A.~Castro$^{5}$, 
   F.~Figueras$^{4}$, F. Fusi Pecci$^{1}$, S. Galleti$^{1}$,
   \newauthor 
   M. Gebran$^{6}$, C. Jordi$^{4}$, C.~Lardo$^{7}$, E.~Masana$^{4}$, 
   M.~Mongui\'o$^{4}$, P.~Montegriffo$^{1}$, 
   \newauthor
   S.~Ragaini$^{1}$, 
   W.~Schuster$^{5}$, S.~Trager$^{8}$, F.~Vilardell$^{9}$, 
   and H.~Voss$^{4}$\thanks{Based on data obtained within the Gaia
   DPAC (Data Processing and Analysis Consortium) --- and coordinated by the
   GBOG (Ground-based Observations for Gaia) working group --- at various
   telescopes; see acknowlegements.}\\
   $^{1}$Osservatorio Astronomico di Bologna, INAF, Via C. Ranzani, 1, 
         I-40127, Bologna, Italy\\
   $^{2}$Osservatorio Astronomico di Roma, INAF, Via di Frascati, 33, 
         I-00040, Monte Porzio Catone, Italy\\
   $^{3}$ASI Science Data Center c/o ESA-ESRIN, Via Galileo Galilei, s/n,
             I-00044, Frascati, Italy\\
   $^{4}$Departament d'Astronomia i Meteorologia,Institut del Ci\`ences del
             Cosmos (ICC), Universitat de Barcelona (IEEC-UB), c/ Mart\'\i\  i
             Franqu\`es, 1, 08028 Barcelona, Spain\\
   $^{5}$Observatorio Astron\'omico Nacional, Universidad Nacional
	     Aut\'onoma de M\'exico, Apartado Postal 877, C.~P.~22800 Ensenada,
	     B.~C., M\'exico\\
   $^{6}$Department of Physics and Astronomy, Notre Dame University-Louiaze,
             PO Box 72, Zouk Mikael, Zouk Mosbeh, Lebanon\\
   $^{7}$Dipartimento di Astronomia, Universit\`a di Bologna, 
              Via C. Ranzani, 1, I-40127 Bologna, Italy\\
   $^{8}$Kapteyn Institute, University of Groningen, P.O. Box 800, 9700 
	     AV Groningen, the Netherlands\\
   $^{9}$Institut d'Estudis Espacial de Catalunya, Edifici Nexus, c/ 
             Capit\'a, 2-4, desp. 201, E-08034 Barcelona, Spain}
\begin{document}

\date{Accepted XXXX December XX. Received XXXX December XX; in original form 
XXXX October XX}

\pagerange{\pageref{firstpage}--\pageref{lastpage}} \pubyear{2012}

\maketitle

\label{firstpage}

\begin{abstract}
  We describe two ground based observing campaigns aimed at building a grid of
  approximately 200 spectrophotometric standard stars (SPSS), with an internal
  $\simeq$1\% precision and tied to Vega within $\simeq$3\%, for the absolute
  flux calibration of data gathered by Gaia, the ESA astrometric mission. The
  criteria for the selection and a list of candidates are presented, together
  with a description of the survey strategy and the adopted data analysis
  methods. We also discuss a short list of notable rejected  SPSS candidates and
  difficult cases, based on identification problems, literature  discordant
  data, visual companions, and variability. In fact, all candidates are  also
  monitored for constancy (within $\pm$5~mmag, approximately). In particular, we
  report on a CALSPEC standard, 1740346, that we found to be a $\delta$~Scuti
  variable during our short-term monitoring (1--2~h) campaign.
\end{abstract}


\section{Introduction}

\begin{figure}
\includegraphics[width=\columnwidth,height=7cm]{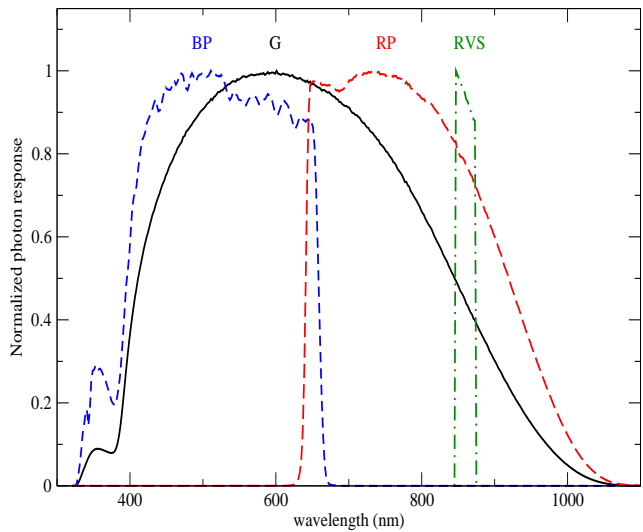}
\caption{The photon response functions of the Gaia G, BP, RP and RVS passbands.
\label{fig_bands}}
\end{figure}

Gaia is an ESA (European Space Agency) all sky astrometric, photometric, and
spectroscopic survey mission aimed at measuring parallaxes, proper motions,
radial velocities, and astrophysical parameters of $\simeq$10$^9$ stars
($\simeq$1\% of the Galactic stellar population) down to magnitude
G$\simeq$20\footnote{the Gaia G-band is the unfiltered broad band defined by the
instrumental response curve, see also Figure~\ref{fig_bands}, extracted from
\citet{jordi10}.}, corresponding to V$\simeq$20--25~mag, depending on spectral
type. 

The astrometric accuracy is expected to be 5--14~$\mu$as for bright stars
(V$<$12~mag), and to reach $\simeq$300~$\mu$as down to V$\simeq$20~mag. Radial
velocities will be measured for stars brighter than V$\simeq$17~mag, depending on
spectral type, and their precision will range from 1~km~s$^{-1}$ for the bright
stars down to 15--20~km~s$^{-1}$ for the faintest stars, bluer stars having
higher uncertainties. The updated science performances of Gaia can be found on
the Gaia ESA
webpage\footnote{http://www.rssd.esa.int/index.php?project=GAIA\&page=index}.

The expected launch will be in August 2013, from the ESA launch site at Kourou
in French Guiana. Gaia will operate for approximately 5 years, with a possible 1
year extension, and the final catalogue is expected to be published 3 years
after mission completion, while a set of intermediate releases is presently
being defined.

Although the primary scientific goal of Gaia is the characterization of the Milky
Way, its scientific impact will range from solar system studies to distant
quasars, from unresolved galaxies to binaries, from supernovae to microlensing
events, from fundamental physics to stellar variability. The wide variety of
scientific topics is illustrated by almost 900 papers in ADS (Astrophysics Data
System, of which more than 200 refereed) to date, on a diversity of subjects,
from the description of various mission components (including software,
pipelines, data treatment philosophy) to simulations of the expected scientific
harvest in many diverse areas. Some papers summarize the expected science results
\citep[see, e.g.,][]{mignard05}, but no single paper can be complete in this
respect, given the huge range of possibilities opened by Gaia.  

Three main instruments can be found on board Gaia, the AF (Astrometric Field),
consisting of 62 CCDs illuminated with white light, which will provide
astrometric measurements and integrated Gaia G-band magnitudes (hereafter G);
the BP (Blue Photometer) and RP (Red Photometer), consisting of two strips of 7
CCDs each and  providing prism dispersed, slitless spectra at a resolution of
R=$\lambda /\delta\lambda$$\simeq$20--100, covering the passbands shown in
Figure~\ref{fig_bands} and also providing integrated BP and RP magnitudes
(hereafter G$_{\rm{BP}}$ and G$_{\rm{RP}}$) and the G$_{\rm{BP}}$--G$_{\rm{RP}}$
colour, which will be fundamental for chromaticity corrections of the
astrometric measurements; and the RVS (Radial Velocity Spectrograph), providing
R$\simeq$11000 spectra in the calcium triplet region (8470--8740~\AA) projected
onto 12 CCDs. The mission output will thus be accurate positions, proper motions
and parallaxes, low resolution BP/RP spectra, integrated G, G$_{\rm{BP}}$, and
G$_{\rm{RP}}$ magnitudes and the G$_{\rm{BP}}$--G$_{\rm{RP}}$ colour, plus
medium resolution RVS spectra and radial velocities for stars brighter than
V$\simeq$17~mag. A classification of all observed objects will be performed on
the basis of BP/RP and RVS spectra and -- when possible -- their parametrization
will be performed as well, which for stars will provide T$_{\rm eff}$, log$g$,
E(B--V), [Fe/H], and [$\alpha$/Fe].

Although Gaia is in principle a self-calibrating mission, some Gaia measurements
need to be tied to existing absolute reference systems, and many Gaia algorithms
need to be trained. Thus extensive theoretical computations and observing
campaigns are being carried out. To make a few examples: radial velocity
standards that are stable to 1~km~s$^{-1}$ are being obtained \citep{crifo10};
extended libraries of observed and theoretical spectra
\citep{PAT-004,sordo10,tsalmantza12}\footnote{\citet{PAT-004}, as many other
documents cited in the following, is a Gaia technical report that is normally not
available to the public. We nevertheless will cite some of these documents
because they contain more detailed discussions of the topics treated here, or
simply to give appropriate credit to work that was done previously. Future papers
of this series will enter in more technical and scientific details. Subject to
approval by the ESA and the Gaia DPAC (Data Processing and Analysis Consortium)
governing bodies, Gaia technical reports can be provided to interested readers by
the authors.} are being established; the Ecliptic poles -- that will be
repeatedly observed in the initial calibration phase of Gaia observations -- are
being observed to produce catalogues of magnitudes and high-resolution spectra
\citep{altman09}. Also, the selection and analysis of reference stars (and
galaxies, quasars, asteroids, solar system objects and so on) for the training of
Gaia algorithms is being carried on by different groups. 

This paper is the first of a series, which will present different aspects of the
survey and of its data products. The series will include technical papers on the
instrumental characterization, data papers presenting flux tables, photometric
measurements, and lightcurves of our SPSS candidates, and scientific follow-up
papers based on survey data and, when needed, on additional data.

This paper presents the ongoing observational survey aimed at building the grid
of spectrophotometric standard stars (SPSS) for the absolute flux calibration of
Gaia spectra and integrated magnitudes. The structure of the paper is the
following: the Gaia external calibration model is briefly illustrated in
Section~\ref{sec-model}; the selection criteria and a list of candidate SPSS are
presented in Section~\ref{sec-spss}; the observing campaigns and facilities are
described in Section~\ref{sec-survey}; a description of the data treatment
principles and methods can be found in Section~\ref{sec-reds}; and a set of
preliminary results is presented in Section~\ref{sec-res}.

\section{Flux calibration model} 
\label{sec-model}

Calibrating (spectro)photometry obtained from the usual type of ground based
observations (broadband imaging, spectroscopy) is not a trivial task, but  the
procedures are well known \citep[see, e.g.,][]{bessell99} and many groups have
developed sets of appropriate standard stars for the more than 200 photometric
known systems, and for spectroscopic observations.

In the case of Gaia, several instrumental effects -- much more complex than those
usually encountered -- redistribute light along the SED (Spectral Energy
Distribution) of the observed objects. The most difficult Gaia data to calibrate
are the BP and RP slitless spectra, requiring a new approach to the derivation of
the calibration model and to the SPSS grid needed to perform the actual
calibration. Some important complicating effects are:

\begin{itemize}
\item{the large focal plane with its large number of CCDs makes it so that
different observations of the same star will be generally on different CCDs, with
different quantum efficiencies, optical distorsions, transmissivity and so on.
Therefore, each wavelength and each position across the focal plane has its
(sometimes very different) PSF (point spread function);} 
\item{TDI (Time Delayed Integration) continuous reading mode, combined with the
need of compressing most of the data before on-ground transmission, make it
necessary to translate the full PSF into a linear (compressed into 1D) LSF (Line
Spread Function), which of course adds complication into the picture;}
\item{in-flight instrument monitoring is foreseen, but never comparable to the full
characterization that will be performed before launch, so the real instrument -- at
a  certain observation time -- will be slightly different from the theoretical one
assumed initially, and this difference will change with time;} 
\item{finally, radiation damage (or CTI, Charge Transfer Inefficiencies)
deserves special mention, for it is one of the most important factors in the
time variation of the instrument model \citep{weiler11,prod'homme11,pasquier11}.
It has particular impact onto the BP and RP dispersed images because the objects
travel along the BP and RP CCD strips in a direction that is parallel to the
spectral dispersion (wavelength coordinate) and therefore the net effect of
radiation damage can be to alter the SED of some spectra. Several solutions are
being implemented to mitigate CTI effects, but the global instrument complexity
calls for a new approach to spectra flux calibrations.}
\end{itemize}

A flux calibration model is currently implemented in the photometric pipeline,
which splits the calibration into an {\em internal} and an {\em external} part.
The internal calibration model \citep{CJ-042,jordi11,JMC-006,CF-012} uses a large
number of well behaved stars (internal standards), observed by Gaia, to report
all observations to a {\em reference} instrument, on the same instrumental
relative flux and wavelength scales. Once each observation for each object is
reported to the internal reference scales, the absolute or {\em external}
calibration \citep{PMN-003,SR-001,SR-002,SR-003} will use an appropriate SPSS set
to report the relative flux scale to an absolute flux scale in physical units,
tied to the calibration of Vega (see also Section~\ref{sec-spss}). Alternative
approaches where the internal and external calibration steps are more
inter-connected are being tested to maximise the precision and the accuracy of
the Gaia calibration \citep{AB-020,PMN-005,PMN-006,JMC-011}. The Gaia calibration
model was also described by \citet{pancino10}, \citet{jordi11}, and
\citet{cacciari11}.

The final flux calibrated products will be: averaged (on all transits -- or
observations) white light magnitudes, G; integrated BP/RP magnitudes,
G$_{\rm{BP}}$ and G$_{\rm{RP}}$; flux calibrated BP/RP spectra; RVS spectra and
integrated G$_{\rm RVS}$ magnitudes, possibly also flux calibrated
\citep{ST-002}. The G$_{\rm{BP}}$--G$_{\rm{RP}}$ colour will be used to correct
for chromaticity effects in the global astrometric solution. Only for specific
classes of objects, epoch spectra and magnitudes will be released, with variable
stars as an obvious example. 

The external calibration model contains -- as discussed -- a large number of
parameters, requiring a large number (about 200) of calibrators. With the
standard calibration techniques \citep{bessell99}, the best possible calibrators
are hot, almost featureless stars such as WD or hot subdwarfs. Unfortunately,
these stars are all similar to each other, forming an intrinsically degenerate
set. The Gaia calibration model instead requires to differentiate as much as
possible the calibrators, by including smooth spectra, but also spectra with
absorption features, both narrow (atomic lines) or wide (molecular bands),
appearing both on the blue and the red side of the spectrum\footnote{Including
emission line objects in our set of calibrators is problematic. Emission line
stars are often variable and thus do not make good calibrators. Similarly for
Quasars, which are typically faint for our ground-based campaigns. Thus, with
this calibration model we do not expect to be able to calibrate with very high
accuracy emission line objects.}. An experiment described by \citet{pancino10}
shows that the inclusion of just a few M stars\footnote{While M giants show
almost always variations of the order of 0.1--0.2~mag, and thus are not useful as
flux standards, M dwarfs rarely do \citep{eyer08}.} with large molecular
absorptions in the Gaia SPSS set can improve the calibration of similarly red
stars by a factor of more than ten (from a formal error of 0.15~mag to an error
smaller than 0.01~mag). 

\begin{figure}[!t]
\centering
\includegraphics[width=\columnwidth]{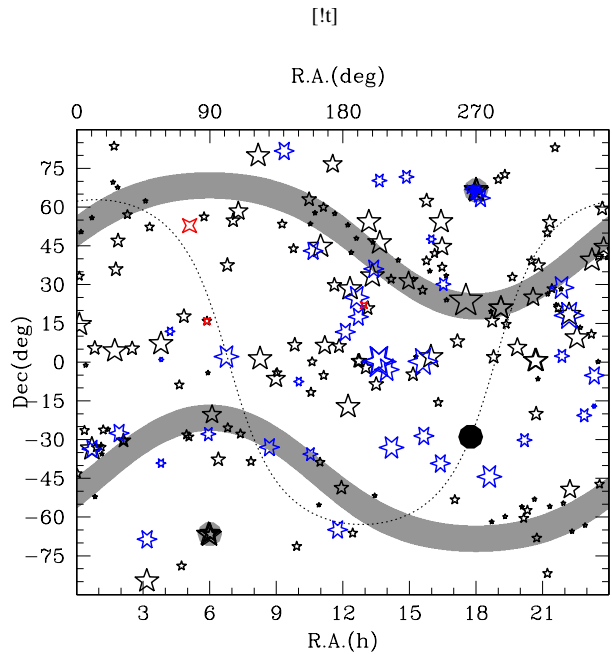}
\caption{Distribution of our SPSS candidates on the sky. The Galactic plane and
center are marked with a dotted line and a large black circle, respectively. The
Ecliptic poles are marked as two large grey circles, and two stripes at
$\pm$45~deg from the Ecliptic poles (roughly where Gaia is observing more often)
are shaded in grey. Our {\em Pillars} are shown as three four-pointed stars, the
{\em Primary SPSS candidates} as six-pointed stars, and the {\em Secondary SPSS
candidates} as five-pointed stars. The stars size is proportional to the SPSS
brightness, ranging from V$\simeq$8 (largest symbols) to 15~mag (smallest
symbols), approximately.}
\label{fig_sky}
\end{figure}

In conclusion, the complexity of the instrument reflects in a complex calibration
model, that requires a large set of homogeneously calibrated SPSS, covering a
range of spectral types. No such database exists in the literature, and new
observations are necessary to build it.

\section{The candidate SPSS}
\label{sec-spss}

The Gaia photometric calibration model implies specific needs as it comes to
{\em (i)} the selection criteria of the SPSS candidates and {\em (ii)} the
characteristics of their flux tables (i.e., their calibrated spectra). The
derived formal requirements \citep{FVL-072} define both the SPSS grid and the
observing needs and can be summarized as follows:

\begin{figure}[!t]
\centering
\includegraphics[width=\columnwidth]{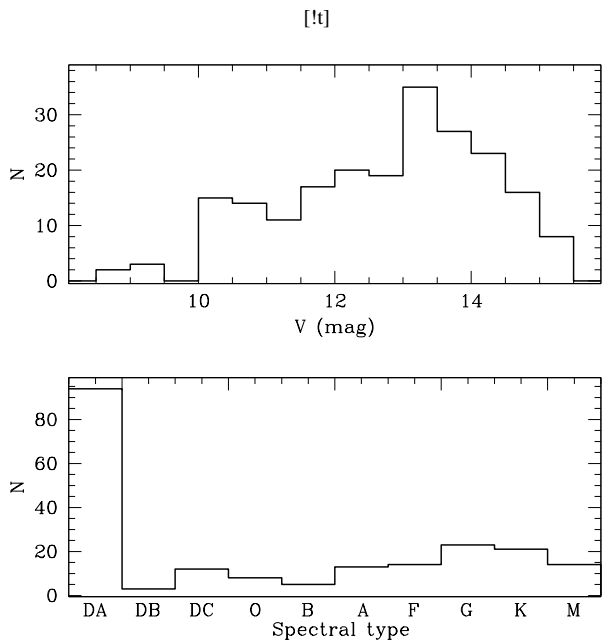} 
\caption{Distribution of all our SPSS candidates in magnitude (top panel) and
spectral type (bottom panel).}
\label{fig_hist} 
\end{figure}

\begin{itemize}
\item{spectral resolution R=$\lambda/\delta\lambda \simeq 1000$, i.e., they
should  oversample the Gaia BP/RP resolution by a factor of  4--5 at least;}
\item{wavelength coverage: 3300--10500~\AA, corresponding to the full coverage of
the BP and RP spectrophotometers;}
\item{large sample (approximately 200--300 stars), covering different spectral
types, although a large fraction should consist in hot stars, as featureless as
possible;}
\item{magnitude range 9$<$V$<$15~mag: when observed by Gaia they should ensure
an end-of-mission S/N$\simeq$100 over most of the wavelength range, without
saturating;}
\item{typical uncertainty on the absolute flux, with respect to the
assumed calibration of Vega \citep{bohlin04,bohlin07}\footnote{A great promise
for the future of flux calibrations comes from the ACCESS mission
\citep{kaiser10}. We tried to include a few of their primary targets in our SPSS
candidates list.} of $\simeq$3\%, excluding small troubled areas in the
spectral range (telluric bands residuals, extreme red and blue edges), where it
can be somewhat worse;}
\item{very homogeneous data treatment and quality, i.e., the SPSS flux tables
should have $\simeq$1\% internal precision;}
\item{photometric stability within $\pm$5~mmag, necessary to ensure the
above accuracy and precision.}
\end{itemize}

The CALSPEC\footnote{http://www.stsci.edu/hst/observatory/cdbs/calspec.html}
\citep{bohlin07} and the \citet{stritzinger05} databases are very good starting
points \citep[see also][for further references]{bessell12}, but new observations
are needed.   

It is clear that if we add the requirements deriving from a ground-based
campaign\footnote{Observations must be feasible with 2--4~m class telescopes, all
year round from both hemispheres, and the SPSS must be free from relatively
bright companions, that might be seen as separate objects from space, but are
close enough to contaminate the SPSS aperture photometry and wide slit spectra,
owing to the Earth's atmospheric seeing.} to the above ones, it becomes very
difficult to assemble the grid in a relatively short time. Therefore we decided
to proceed in steps. The link between Vega and our SPSS will be ensured by three
{\em Pillars} (Section~\ref{sec-pillars}); these will enable to calibrate the
{\em Primary SPSS} (Section~\ref{sec-primaries}), our ground-based calibrators
spread over the whole sky. The primary SPSS will in turn enable to calibrate our
{\em Secondary SPSS} (Section~\ref{sec-secondaries}), which constitute the actual
Gaia grid, together with the eligible {\em Primaries}. The basic principles of
our calibration strategy were first outlined by \citet{MBZ-001}. The sky
distribution of our candidates is shown in Figure~\ref{fig_sky}, while the
magnitude and spectral type distributions are shown in Figure~\ref{fig_hist}. 
More details on the selection criteria, sources, and candidate lists can be found
in \citet{GA-001} and \citet{GA-003}.

\subsection{Pillars}
\label{sec-pillars}

Our three pillars are the CALSPEC pillars and were selected from \citet{bohlin95}
and \citet{bohlin96}. They are the DA (pure hydrogen atmosphere) white dwarfs
(WD) named G~191-B2B, GD~71, and GD~153, three well known and widely used
standards. A fourth star from \citet{bohlin95}, HZ~43, was excluded from our list
because it is member of a binary system. Its companion, a dMe star
\citep{dupuis98}, at a distance of  $\simeq$3$"$, is brighter longward of
$\simeq$7000\AA\ \citep{bohlin01}, and therefore not usable in our ground-based
campaign, where the actual seeing ranges from $\simeq$0.5" up to $>$2" in some
cases and the slit width is 10"-12" for our spectra.

The flux calibrated spectra of the {\em Pillars}, available in the CALSPEC
database, are tied to the revised Vega flux\footnote{Vega was calibrated using
STIS (Space Telescope Imaging Spectrograph) observations \citep{bohlin04} and
the calibration was later revised by \citet{bohlin07}.}, and their flux
calibrations are based on the comparison of WD model atmospheres\footnote{Hubeny
NLTE models \citep{hubeny95}. See also \cite{bohlin07} and references therein.
In particular, these model flux distributions are normalized to an absolute flux
of Vega of $3.46\times10^{-9}$erg cm$^{-2}$ s$^{-1}$ \AA$^{-1}$ at 5556~\AA.}
and spectra obtained with the Faint Object Spectrograph (FOS) aboard HST. The
{\em Pillars} are in the temperature range  $32\,000 \leq  T_{\rm{eff}} \leq
61\,000$~K and the FOS spectrophotometry agrees with the model fluxes to within
2\% over the whole UV-visible range. In addition, the simulated B and V
magnitudes of the data agree to better than 1\% with the Landolt photometry
\citep{landolt07}. 

Some of the most recent literature measurements for the three {\em Pillars} are
listed in Table~\ref{tab_pillars}. 

\begin{table}
\centering
\begin{minipage}{85mm}
\caption{Pillars\label{tab_pillars}}
\begin{tabular}{@{\extracolsep{-7pt}}lccccl}
\hline
Star & RA (J2000)\footnote{\citet{vanleeuwen07} coordinates;
        $^{b}$ \citet{bohlin05} magnitudes and spectral types;
        $^{c}$ \citet{landolt07} magnitudes.} 
     & Dec (J2000)$^{a}$ & B & V 
     & Type$^{b}$ \\
     & (hh:mm:ss) & (dd:pp:ss) & (mag) & (mag) & \\
\hline
G~191-B2B& 05:05:30.61 & +52:49:51.95 & 11.46$^{c}$ & 11.78$^{c}$ & DA0 \\
GD~71    & 05:52:27.63 & +15:53:13.37 & 12.78$^{b}$ & 13.03$^{b}$ & DA1 \\
GD~153   & 12:57:02.33 & +22:01:52.52 & 13.07$^{b}$ & 13.35$^{b}$ & DA1 \\
\hline
\end{tabular}
\end{minipage}
\end{table}

\subsection{Primary SPSS candidates}
\label{sec-primaries}

The candidate {\em Primary} SPSS are 44 bright (9$\la$V$\la$14~mag --- see also
Table~\ref{tab_primaries}), well known spectrophotometric  standards with spectra
already in the CALSPEC flux scale, or which can be easily tied to that scale with
dedicated ground-based observations. We selected them according to the criteria
outlined above, and with the additional criterium that the sample should be
observable from both hemispheres, all year round, with 2--4~m class telescopes,
as mentioned above.  

We searched for candidates the best existing datasets, such as CALSPEC,
\citet{oke90}, \citet{hamuy92,hamuy94}, \citet{stritzinger05}. As already noted,
the {\em  Primary} SPSS will be calibrated using the {\em Pillars}, and will
constitute our grid of ground-based calibrators for the {\em Secondary} SPSS.
Those {\em Primaries} which {\em a posteriori} will satisfy also the criteria
outlined for the {\em Secondary} SPSS (e.g., will have an end-of-mission
satisfactory S/N ratio when observed by Gaia) will be included in the final list
of Gaia SPSS. The {\em Primary} SPSS candidates are listed in
Table~\ref{tab_primaries} along with some recent literature information.

We mention here that one of the CALSPEC standards, star 1740346, was found to be
a variable with an amplitude of the order of 10 mmag, and is probably a
$\delta$~Scuti type variable, as described in Section~\ref{sec-var}. We are
gathering additional data to characterize its variability. 

\begin{table}
\begin{minipage}{85mm}
\caption{Primary SPSS candidates\label{tab_primaries}}
\small
\begin{tabular}{@{\extracolsep{-11pt}}lccccl}
\hline
Star & RA (J2000) & Dec (J2000)	& B & V & Type \\
     & (hh:mm:ss) & (dd:pp:ss) & (mag) & (mag) & \\
\hline
\noalign{\smallskip} 
\multicolumn{6}{c}{\em White dwarfs and hot subdwarfs:}\\
EG~21       & 03:10:31.02\footnote{\citet{perryman97a}; $^{b}$\citet{hamuy92}; 
                $^{c}$\citet{bica96}; $^{d}$\citet{hawarden01}; $^{e}$\citet{landolt92};
                $^{f}$2MASS \citep{cutri03}; $^{g}$\citet{landolt07}; $^{h}$\citet{bakos02};
                $^{i}$\citet{bessell99}; $^{j}$\citet{turnshek90}; $^{k}$\citet{ostensen10};
                $^{l}$\citet{colina94}; $^{m}$\citet{holberg02}; $^{n}$\citet{pokorny03};
                $^{o}$\citet{stone83}; $^{p}$\citet{hog98}; $^{q}$\citet{stone77};
                $^{r}$\citet{stritzinger05}; $^{s}$\citet{landolt83}; 
                $^{t}$\citet{sembach92}; $^{u}$\citet{kilkenny89}; $^{v}$\citet{salim03};
                $^{w}$\citet{bohlin11}; $^{x}$\citet{reach05}; $^{y}$This star was wrongly 
                identified by \citet{hamuy92} as LTT~377; the case is discussed in detail in 
                Section~\ref{sec-id}; $^{z}$\citet{drilling79}; $^{aa}$\citet{zacharias09};
                $^{bb}$\citet{hog00}; $^{cc}$\citet{bohlin05}; $^{dd}$\citet{casagrande06}; 
                $^{ee}$\citet{koen10}; $^{ff}$\citet{sion09}; $^{gg}$\citet{hog98}, for 
                approximate Johnson magnitudes the formulae V=VT--0.090*(BT--VT) and 
                B--V=0.850*(BT--VT) where used; $^{hh}$Henry Draper Catalogue \citep{cannon93}.}
                                 & --68:36:03.39$^{a}$  & 11.42$^{b}$  & 11.38$^{b}$ & DA3$^{c}$ \\
GD~50       & 03:48:50.20$^{d}$  & --00:58:31.20$^{d}$  & 13.79$^{e}$  & 14.06$^{e}$ & DA2$^{c}$ \\
HZ~2        & 04:12:43.55$^{f}$  &  +11:51:49.00$^{f}$  & 13.79$^{g}$  & 13.88$^{g}$ & DA3$^{c}$ \\
LTT~3218    & 08:41:32.56$^{h}$  & --32:56:34.90$^{h}$  & 12.07$^{e}$  & 11.85$^{e}$ & DA$^{i}$  \\
AGK+81266   & 09:21:19.18$^{a}$  &  +81:43:27.64$^{a}$  & 11.60$^{g}$  & 11.94$^{g}$ & O$^{j}$   \\
GD~108      & 10:00:47.37$^{k}$  & --07:33:30.50$^{k}$  & 13.34$^{l}$  & 13.58$^{l}$ & B$^{k}$   \\
Feige~34    & 10:39:36.74$^{a}$  &  +43:06:09.25$^{a}$  & 10.84$^{g}$  & 11.18$^{g}$ & DO$^{j}$  \\
LTT~4364    & 11:45:42.92$^{a}$  & --64:50:29.46$^{a}$  & 11.69$^{e}$  & 11.50$^{e}$ & DQ6$^{i}$ \\
Feige~66    & 12:37:23.52$^{a}$  &  +25:03:59.87$^{a}$  & 10.22$^{g}$  & 10.51$^{g}$ & O$^{j}$   \\
Feige~67    & 12:41:51.79$^{a}$  &  +17:31:19.75$^{a}$  & 11.48$^{g}$  & 11.82$^{g}$ & O$^{j}$   \\
HZ~44       & 13:23:35.26$^{a}$  &  +36:07:59.51$^{a}$  & 11.38$^{g}$  & 11.67$^{g}$ & O$^{j}$   \\
GRW+705824  & 13:38:50.47$^{a}$  &  +70:17:07.62$^{a}$  & 12.68$^{g}$  & 12.77$^{g}$ & DA3$^{j}$ \\
EG~274      & 16:23:33.84$^{a}$  & --39:13:46.16$^{a}$  & 10.89$^{b}$  & 11.02$^{b}$ & DA2$^{c}$ \\
EG~131      & 19:20:34.93$^{a}$  & --07:40:00.05$^{a}$  & 12.35$^{ee}$ & 12.29$^{ee}$& DBQA5$^{}$\\
LTT~7987    & 20:10:56.85$^{a}$  & --30:13:06.64$^{a}$  & 12.27$^{e}$  & 12.21$^{e}$ & DA4$^{m}$ \\
G~93-48     & 21:52:25.38$^{a}$  &  +02:23:19.56$^{a}$  & 12.73$^{e}$  & 12.74$^{e}$ & DA3$^{c}$ \\ 
LTT~9491    & 23:19:35.44$^{n}$  & --17:05:28.40$^{n}$  & 14.13$^{g}$  & 14.11$^{g}$ & DB3$^{i}$ \\
Feige~110   & 23:19:58.40$^{a}$  & --05:09:56.21$^{a}$  & 11.53$^{g}$  & 11.83$^{g}$ & O$^{o}$   \\ 
\hline
\noalign{\smallskip} 
\multicolumn{6}{c}{\em Other hot stars (O, B, and A):}\\
HD 37725    & 05:41:54.37$^{p}$  &  +29:17:50.93$^{p}$  &  8.12$^{gg}$ &  8.31$^{gg}$& A3$^{hh}$ \\
HILT~600    & 06:45:13.37$^{p}$  &  +02:08:14.70$^{p}$  & 10.62$^{b}$  & 10.44$^{b}$ & B1$^{q}$  \\
Feige~56    & 12:06:47.23$^{a}$  &  +11:40:12.64$^{a}$  & 10.93$^{b}$  & 11.06$^{b}$ & B5p$^{b}$ \\
SA~105-448  & 13:37:47.07$^{p}$  & --00:37:33.02$^{p}$  &  9.44$^{r}$  &  9.19$^{r}$ & A3$^{s}$  \\    
HD~121968   & 13:58:51.17$^{a}$  & --02:54:52.32$^{a}$  & 10.08$^{r}$  & 10.26$^{r}$ & B1$^{t}$  \\  
CD-32~9927  & 14:11:46.32$^{p}$  & --33:03:14.30$^{p}$  & 10.84$^{u}$  & 10.44$^{u}$ & A0$^{o}$  \\
LTT~6248    & 15 38 59.66$^{v}$  & --28 35 36.87$^{v}$  & 12.29$^{e}$  & 11.80$^{e}$ & A$^{b}$   \\
1743045     & 17:43:04.48$^{f}$  &  +66:55:01.60$^{f}$  & 13.80$^{w}$  & 13.52$^{w}$ & A5$^{x}$  \\
1805292     & 18:05:29.28$^{f}$  &  +64:27:52.00$^{f}$  & 12.50$^{w}$  & 12.06$^{w}$ & A6$^{w}$  \\
1812095     & 18:12:09.57$^{f}$  &  +63:29:42.30$^{f}$  & 11.90$^{w}$  & 11.80$^{w}$ & A5$^{w}$  \\
BD+28~4211  & 21:51:11.02$^{a}$  &  +28:51:50:36$^{a}$  & 10.17$^{g}$  & 10.51$^{g}$ & Op$^{j}$  \\
\hline
\noalign{\smallskip} 
\multicolumn{6}{c}{\em Cooler stars (F, G, and K):}\\
CD-34~241$^{y}$
            & 00:41:46.92$^{p}$  & --33:39:08.51$^{p}$  & 11.71$^{b}$  & 11.23$^{b}$  & F$^{b}$  \\
LTT~1020    & 01:54:50.27$^{v}$  & --27:28:35.74$^{v}$  & 12.06$^{e}$  & 11.51$^{e}$  & G$^{b}$  \\
LTT~1788    & 03:48:22.67$^{n}$  &  -39:08:37.20$^{n}$  & 13.61$^{e}$  & 13.15$^{e}$  & F$^{b}$  \\
LTT~2415    & 05:56:24.74$^{a}$  & --27:51:32.35$^{a}$  & 12.60$^{e}$  & 12.20$^{e}$  & G$^{i}$  \\
LTT~3864    & 10:32:13.60$^{v}$  & --35:37:41.80$^{v}$  & 12.65$^{e}$  & 12.17$^{e}$  & F$^{b}$  \\ 
SA~105-663  & 13:37:30.34$^{a}$  & --00:13:17.37$^{a}$  &  9.10$^{s}$  &  8.76$^{s}$  & F$^{z}$  \\       
P~41-C      & 14:51:57.99$^{aa}$ &  +71:43:17.38$^{aa}$ & 12.84$^{bb}$ & 12.16$^{bb}$ & G0$^{cc}$\\
SA~107-544  & 15:36:48.10$^{p}$  & --00:15:07.11$^{p}$  &  9.44$^{s}$  &  9.04$^{s}$  & F3$^{z}$ \\
P~177-D     & 15:59:13.57$^{f}$  &  +47:36:41.90$^{f}$  & 13.96$^{dd}$ & 13.36$^{dd}$ & G0$^{cc}$\\
P~330-E     & 16:31:33.82$^{f}$  &  +30:08:46.50$^{f}$  & 13.52$^{dd}$ & 12.92$^{dd}$ & G0$^{cc}$\\
KF08T3      & 17:55:16.23$^{f}$  &  +66:10:11.70$^{f}$  & 14.30$^{cc}$ & 13.50$^{cc}$ & K0$^{x}$ \\
KF06T1      & 17:57:58.49$^{f}$  &  +66:52:29.40$^{f}$  & 14.50$^{cc}$ & 13.52$^{cc}$ & K1$^{x}$ \\
KF06T2      & 17:58:37.99$^{f}$  &  +66:46:52.20$^{f}$  & 15.10$^{cc}$ & 13.80$^{cc}$ & K1$^{x}$ \\
KF01T5      & 18:04:03.80$^{x}$  &  +66:55:43.00$^{x}$  & ...          & 13.56$^{x}$  & K1$^{x}$ \\
LTT~7379    & 18:36:25.95$^{a}$  & --44:18:36.94$^{a}$  & 10.83$^{e}$  & 10.22$^{e}$  & G0$^{b}$ \\
BD+17~4708  & 22:11:31.37$^{a}$  &  +18:05:34.17$^{a}$  &  9.91$^{g}$  &  9.46$^{g}$  & F8$^{cc}$\\
LTT~9239    & 22:52:41.03$^{v}$  & --20:35:32.89$^{v}$  & 12.67$^{e}$  & 12.07$^{e}$  & F$^{b}$  \\ 
\hline
\end{tabular}
\end{minipage}
\end{table}

\subsection{Secondary SPSS candidates}
\label{sec-secondaries}

The {\em Secondary} SPSS are selected according to the criteria given above; in
particular they need to provide BP/RP spectra with an adequate end-of-mission S/N
ratio (see above). This was statistically verified for all our SPSS candidates
\citep{JMC-001,JMC-002} with a set of simulations of the expected number of
transits depending on the position on the sky and on the launch conditions. Stars
fainter than V$\simeq$13~mag need to have a higher number of transits to gather
sufficient end-of-mission S/N when observed by Gaia. The candidates surviving
this test are presented in Table~\ref{tab_secondaries} along with some recent
literature information. Not all literature data (especially magnitudes and
spectral types) have the same precision\footnote{Literature data come from a
variety of heterogeneous sources, and are determined with many diferent methods.
In particular, in Tables~\ref{tab_primaries} and \ref{tab_secondaries}, the most
uncertain magnitudes are those derived with the approximated formulae from the
TYCHO magnitudes \citep{hog98}, while the most uncertain spectral types are the
ones roughly estimated by us from the \citet{carney94} temperatures.}, but we
gathered the best data available, to our knowledge; we will hopefully produce
more precise information from our own data and, later, from Gaia. Our source
catalogues were mainly (but not only):

\begin{itemize}
\item{the {\em ``Catalog of Spectroscopically Identified White Dwarfs"}
\citep{mccook99}, containing 2249 stars in the original paper, and 12876 in the
online --- regularly updated ---
catalogue\footnote{http://www.astronomy.villanova.edu/WDCatalog/index.html} 
at the time of writing;}
\item{{\em ``A Catalog of Spectroscopically Confirmed White Dwarfs from the
Sloan  Digital Sky Survey Data Release 4" (SDSS)} \citep{eisenstein06},
containing 9316 objects. The complete data-set is available online\footnote{
http://iopscience.iop.org/0067-0049/167/1/40/datafile1.txt};}
\item{a list of 121 DA white dwarfs for which there are
FUSE\footnote{http://fuse.pha.jhu.edu/} data (Barstow 2010, private
communication);}
\item{a selection of metal poor stars from {\em ``A survey of proper motion
stars. 12: an expanded  sample"} \citep{carney94} containing 52 stars (Korn 2010,
private  communication);}
\item{{\em ``The HST/STIS Next Generation Spectral Library"}
\citep[NGSL,][]{gregg04}\footnote{http://archive.stsci.edu/prepds/stisngsl/}
containing 378 bright stars covering a wide range in abundance, effective 
temperature and luminosity;}
\item{the catalogues from {\em ``The M dwarf planet search programme  at the ESO
VLT + UVES. A search for terrestrial planets in the habitable zone  of M dwarfs"}
\citep{zechmeister09} and from  {\em ``Rotational Velocities for M Dwarfs"}
\citep{jenkins09}, particularly useful for the selection of red stars;}
\item{the {\em ``Medium-resolution Isaac Newton Telescope Library of Empirical
Spectra (MILES)''}\footnote{http://www.ucm.es/info/Astrof/miles/miles.html}
\citep{sanchez-blazquez06} database containing 985 spectra obtained at the 2.5~m
Isaac Newton Telescope (INT) covering the range 3525--7500~\AA;}
\item{{\em``SEGUE:  A Spectroscopic Survey of 240,000 Stars with g=14--20''}
\citep{yanny09}, containing $\simeq$240\,000 moderate-resolution spectra from
3900 to 9000~\AA\  of fainter Milky Way stars ($14.0 \leq g \leq 20.3$) of a wide
variety of spectral types and classes. In particular, we made use of the
re-analysis by \citet{PAT-004} and \citet{tsalmantza12} to select a few suitably
bright stars;}
\item{{\em``The Ecliptic Poles Catalogue Version 1.1''}
\citep{altman09}, a preliminary version of the photometric catalogue that will be
used by Gaia in the initial observation phases, containing 150\,000 stars down to 
V$\simeq$22~mag, in a region of approximately 1~deg$^2$ around the Northern and 
Southern Ecliptic poles;}
\item{The WD online catalogue maintained by A.
Kawka\footnote{http://sunstel.asu.cas.cz/$\sim$kawka/Mainbase.html}, and information
from \citet{kawka07};}
\item{A provisional list of targets for the ACCESS mission \citep{kaiser07,kaiser10},
provided by M.~E. Kaiser (2010, private communication).}
\end{itemize}

Other references can be found in Table~\ref{tab_secondaries}. All the lists were
merged and cross-checked to eliminate redundant entries. The clean list
($\simeq$13\,500 stars) was then used to extract a subsample ($\simeq$300 stars)
according to the criteria outlined above. 

During the course of the survey, we rejected a few of the original $\simeq$300
candidates because they were found to be binaries, variables, or they showed
close companions on the basis of our literature monitoring and/or of our data.
The rejection procedure, along with a few interesting cases, is described in
Section~\ref{sec-rej}. A few more candidates may be rejected during the course
of the campaign, and some candidates might be added if needed by the Gaia
photometric pipeline, once it is running on real data.

\section{The survey}
\label{sec-survey}

Our survey is split into two campaigns, the {\em main campaign} dedicated to
obtaining spectrophotometry of all our candidate SPSS, and the {\em auxiliary
campaign} dedicated to monitoring the constancy of our SPSS on relevant
timescales.

\subsection{Main campaign} 
\label{sec-main}

Classical spectrophotometry \citep{bessell99} would clearly be the best approach
to obtain absolutely calibrated flux spectra if we had a dedicated telescope.
However, a pure spectrophotometric approach would require too much time, given
that we need high S/N of 300 stars, in photometric sky conditions, which are rare
except maybe in a few sites. We thus decided for a combined approach
\citep{MBZ-001}, in which spectra are obtained even if the sky is
non-photometric\footnote{The cloud coverage must produce grey extinction
variations, i.e., the extinction must not alter significantly the spectral shape.
This condition is almost always verified in the case of veils or thin clouds
\citep{oke90,pakstiene03}, and can be checked a posteriori for each observing
night.}, providing the correct spectral shape of our SPSS (what we will call {\em
``relative flux calibration"}). Then, imaging in photometric conditions and in
three bands (generally B, V, and R, but sometimes also I and, more rarely, U) is
obtained and calibrated magnitudes are used to scale the spectra to the correct
zeropoint ({\em ``absolute flux calibration by comparison"}). 

The calibrated magnitudes of SPSS will be obtained through {\em at least} three
independent observations in photometric conditions. Our sample contains some
photometric standards from \citet{landolt92}, \citet{landolt07}, and a few
secondary Stetson
standards\footnote{http://www4.cadc-ccda.hia-iha.nrc-cnrc.gc.ca/community/STETSON/}
\citep[see][and online updates]{stetson00}. By comparing the obtained magnitudes
and synthetic magnitudes derived from the relatively calibrated spectra, we can
obtain the necessary zeropoint corrections to correct our spectral flux
calibration. To those spectra obtained in photometric conditions (at the moment
approximately 20--25\% of the total) we will apply the classical method, and this
control sample will allow us to check the validity of the combined spectroscopy
plus photometry approach. 

\subsection{Constancy monitoring}
\label{sec-aux}

This kind of monitoring is necessary for a few reasons. Even stars used for years
as spectrophotometric standards were found to vary when dedicated studies have
been performed  \citep[see e.g., G24-9, that was found to be an eclipsing binary
by][]{landolt07}, and even stars that are apparently safe may show unexpected
variations. Our own survey has already found a few variables and suspected
variables, including one of the CALSPEC standards (Section~\ref{sec-var}). 

White dwarfs may show variability with (multi-)periods from about 1 to 20~min
and amplitudes from  about 1-2\% up to 30\%, i.e., ZZ~Ceti type variability. We
have tried to exclude stars within the instability strips for DAV
\citep{castanheira07}, DBV, and DOV but in many cases the existing information
was not sufficient (or sufficiently accurate) to firmly establish the constant
nature of a given WD. Hence, many of our candidate SPSS needed to be monitored
for constancy on {\em short timescales}, of the order of 1--2~h. Similar
considerations are valid for hot subdwarfs \citep{kilkenny07}. 

Also redder stars are often found to be variable: for example K stars have shown
variability of 5-10\% with periods of the order of days to tens of days
\citep{eyer97}. In addition, binary systems are frequent and eclipsing binaries
can be found at all spectral types. Their periods can span a range from a few
hours to hundreds of days, most of them having $P\simeq$ 1-10~days, 
\citep{dvorak04}. Thus, in addition to our short term monitoring, we are
observing all our SPSS on {\em longer timescales}, of about 3~yrs, with a random
phase sampling approximately 4 times a year, which should be enough to detect
variability, although not for a proper characterization of these newly discovered
variables. Unlike the short-term monitoring, the long-term monitoring can be
picked up by Gaia once it starts operations. Gaia data will help in the
characterization and parametrization of the detected variables by providing, on
average, $\simeq$80 sets of spectra and integrated magnitudes in its 5 years of
operation.

We use relative photometry measurements, with respect to field stars, for both
our short (1--2~h) and long (3~yrs) term monitoring campaigns, aiming at
excluding all stars with a variability larger than $\pm$5~mmag, approximately.
Obviously, as soon as a target is recognized as variable, it is excluded from our
candidate list, but we are aware that some characterization of the variability is
of scientific value, so whenever possible, we follow-up our new variable stars
with imaging, more detailed lightcurves and, when necessary,  spectroscopy.

\subsection{Observing facilities and status}
\label{sec-fac}

We have considered a long list of available facilities in both hemispheres
\citep{LF-001,GA-002}. The eligible instruments must be capable of obtaining low
resolution spectroscopy -- with the characteristics described in
Section~\ref{sec-spss} -- and Johnson-Cousins photometry. At least one site in the
North and one in the South with a high probability of having photometric sky
conditions were necessary. We eventually selected six facilities:

\begin{itemize}
\item{EFOSC2@NTT at the ESO La Silla Observatory, Chile, our Southern
facility for spectroscopy and absolute photometry, and for some constancy
monitoring;}
\item{ROSS@REM at the ESO La Silla Observatory, Chile, our Southern
facility for relative photometry;}
\item{CAFOS@2.2m at the Calar Alto Observatory, Spain, one of our Northern
spectrographs and imagers, for absolute and relative (spectro)photometry;}
\item{DOLoRES@TNG at the Roque de Los Muchachos in La Palma, Spain, one of our
Northern spectrographs and imagers, for absolute and relative (spectro)photometry;}
\item{LaRuca@1.5m at the San Pedro M\'artir Observatory, Mexico, our Northern
source of absolute and relative photometry;}
\item{BFOSC@Cassini in Loiano, Italy, providing a few spectra and more
relative photometry in the Northern hemisphere.}
\end{itemize}

Given the diversity of instruments and observing conditions, we enforced a set of
strict observing protocols \citep{EP-001,EP-003,EP-006}, concerning all aspects of
the photometry and spectroscopy observations, including requirements about the
calibration strategy, and on-the-fly quality control of data acquired at the
telescope (see also Sections~\ref{sec-main} and \ref{sec-aux}). Observations
started in 2007. At the time of writing, the survey has been awarded more than 400
nights of observing time, both in visitor and service mode, of which 25--35\% was
lost due to bad weather or technical reasons, or was of non-optimal quality. The
main campaign should be completed within 2012, with the last ESO run assigned in
July 2012 and the last Calar Alto run in May 2012. The short-term variability
monitoring is 85\% complete and the long-term monitoring will take more time and
will probably be completed around 2013--2014. 
\section{Data treatment and data products}
\label{sec-reds}

The required precision and accuracy of the SPSS calibration imposes the adoption
of strict protocols of instrument characterization, data reduction, quality
control, and data analysis. We will briefly outline below our data treatment
methods, while more details will be published in future papers of the series,
presenting our data products. At the time of writing, reductions are ongoing:
pre-reduction of the obtained data is more than 50\% complete, while the analysis
is advancing for short-term (1--2~h) constancy monitoring ($\simeq$35\% complete)
and less complete for spectroscopy ($\simeq$20\%), absolute photometry (just
started). Long-term (3~yrs) constancy monitoring observations are still
incomplete. 

\subsection{Familiarization plans}

We obtained our data from a variety of instruments, that also were upgraded or
modified during the course of the observations, for example a few CCDs were
substituted by new and better CCDs. A strict characterization of the used
instruments was needed, requiring additional calibration data, taken during
daytime, twilight, and also nighttime. We called these technical projects
``familiarization plans" \citep{GA-004,SMR-002}. Their results will be published
in subsequent technical papers of this series, and they can be roughly summarized
as follows:

\begin{itemize}
\item{CCD familiarization plan, containing a study of the dark and bias frames
stability; the shutter characterization (shutter times and delays); and the study
of the linearity of all employed CCDs;}
\item{Instrument familiarization plan studying the stability of
imaging and spectroscopy flats, the study of fringing, and the lamp flexures of
the employed spectrographs;}
\item{Site familiarization plan (in preparation), providing extinction curves,
extinction coefficients, colour terms, and a study of the effect of
``calima"\footnote{Calima is a dust wind originating in the Sahara air layer,
which often affects observations in the Canary Islands.} on the spectral shape.}
\end{itemize}

As a results of these studies, specific recommendations for observations and data
treatment were defined.

\subsection{Pre-reductions}

Data reductions were performed mostly with IRAF\footnote{IRAF is the Image
Reduction and Analysis Facility, a general purpose software system for the
reduction and analysis of astronomical data. IRAF is written and supported by
the IRAF programming group at the National Optical Astronomy Observatories
(NOAO) in Tucson, Arizona. NOAO is operated by the Association of Universities
for Research in Astronomy (AURA), Inc. under cooperative agreement with the
National Science Foundation} and IRAF-based pipelines. The detailed data
reduction protocols are described in Gaia technical reports
\citep{SMR-001,SMR-003,GCC-001,GA-006}.

\begin{figure}
\includegraphics[width=\columnwidth]{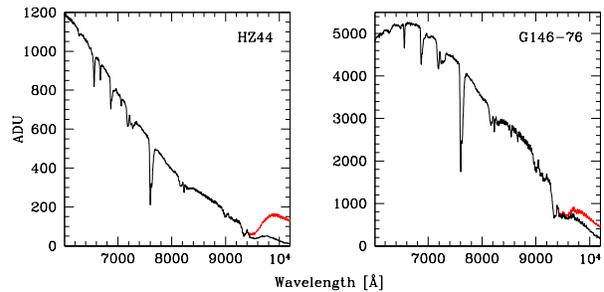}
\caption{Second-order contamination on DOLoRes@TNG spectra of a blue star (left
panel) and a red star (right panel); the black lines are the corrected spectra,
while the red lines above, starting at about 9500~\AA, show the contaminated
spectra.}
\label{fig_2nd}
\end{figure}

For imaging, we pre-reduced the frames with standard techniques, and then
performed aperture photometry with SExtractor \citep{bertin96}. SExtractor also
provides many useful parameters that we will use for a semi-automated quality
control (QC) of each reduced frame, allowing to identify saturated or too faint
SPSS, or frames that do not contain enough good reference stars in the field to
perform relative photometry. Reduced frames that pass QC and their respective
photometric catalogues are stored in our local archive. 

Spectroscopic reductions are less automated, relying mostly on the standard IRAF
longslit package and tasks. Spectra are pre-reduced, extracted and wavelength
calibrated. Spectrophotometry is obtained with a wide slit (5--6 times the
seeing, at least; generally the widest available slits are 10" or 12"). Narrow
slit spectra are also observed (typically with a slit of 1" to 2.5", depending on
the instrument), to obtain a better wavelength calibration. In some cases (slit
larger than 1.5 times the seeing), we will attempt to correct the narrow slit
spectra for differential light losses; tests show that this can be done in most
cases with a third order polynomial fit. The corrected narrow slit spectra will
thus add to the S/N of wide slit spectra, and will also help in beating down the
fringing, because the fringing patterns of wide and narrow slit spectra are
different\footnote{The fringing pattern in the extracted spectra is a combined 1D
result of a 2D pattern, in an aperture that covers a different CCD region in the
wide and in the narrow slit spectra. Thus, the 1D fringing pattern of these two
kinds of spectra will be different.}. Extracted and wavelength calibrated 1D
spectra are stored locally for future processing, if they pass some basic QC (not
saturated or too faint, no close companions in the slit, and so on).

\subsection{Higher level spectra treatment}

After spectra are extracted and wavelength calibrated, they are corrected for
telluric absorption features and for second-order contamination (see below). The
blue and red spectral ranges, that are observed separately with the available
instruments, are joined after performing a relative calibration using the
available {\em Pillar} or {\em Primary} observations taken on the same night at
different airmasses. 

To illustrate the quality of the reduction procedures, we show in
Figure~\ref{fig_2nd} our second-order contamination correction for a blue and a
red star. The effect arises when light from blue wavelengths, from the second
dispersed order of a particular grism or grating, falls on the red wavelengths of
the first dispersed order. Such contamination usually happens when the instrument
has no cross-disperser. Of the instruments we use (Section~\ref{sec-fac}), only
EFOSC2@NTT and DOLoRes@TNG present significant contamination. To map the blue
light falling onto our red spectra, we adapted a method proposed by
\citet{sanchez-blazquez06} and applied it to dedicated observations
\citep{GA-005}. Our wavelength maps generally allow us to recover the correct
spectral shape to within a few percent, as tested on a few CALSPEC standards
observed with both TNG and NTT. 

If the spectra were observed in photometric conditions, after the above
manipulations the flux calibration is complete and ready to be checked. Otherwise,
the shape of the spectrum is recovered, but an additional zeropoint correction is
required. Different levels of intermediate data products are stored after basic
QC, including spectra with and without telluric correction or second-order
contamination correction. 

\subsection{Absolute and relative photometry}

Photometry observations are taken in the form of a night point (absolute or
relative, depending on sky conditions) or a time series. The night point is a
triplet of images in each of three filters (B, V, R, and sometimes also I or U)
taken consecutively. A series lasts at least one hour, contains at least 30
exposures, and is taken with the bluest available filter (B in most cases,
except for REM, where we use V). The SExtractor catalogues are cross-matched
with CataXcorr\footnote{CataXcorr is part of a package dedicated to catalogue
cross-matching and astrometry, developed by P.~Montegriffo at the Bologna
Observatory (INAF).} to identify the SPSS and the reliable reference stars in
the surrounding field. 

Absolute photometry is then performed in a standard way, using observations of
two or three standard fields \citep{landolt92} at different airmasses during the
night. Observations of the same SPSS are taken repeatedly at different times and,
when possible, different sites, to be able to identify any hidden systematics.
Some stars in our candidates list are spectrophotometric standards
\citep{landolt92,landolt07,stetson00} that will be used to check the quality of
our measurements. The final calibrated magnitudes will be used to correct the
zero-point of spectra observed in non-photometric (but grey absorption) sky
conditions, as explained later.

Relative photometry is performed using the difference between the SPSS and the
available field stars (at least two are required) magnitudes. Reference stars must
be non saturated, not too faint, present in all frames, and non variable. Some
preliminary results of this procedure are discussed in Section~\ref{sec-res}. The
target precision of at least 10~mmag, necessary to meet our calibration
requirements (Section~\ref{sec-spss}), is generally always reached with BFOSC,
EFOSC2, LaRuca, DOLoRes, and CAFOS, and most of the times also withROSS@REM, the
robotic telescope in La Silla. 

The final data products of the photometry procedure are absolute magnitudes and
differential lightcurves (on 1--2~h and 3~yr timescales) with their respective
uncertainties.

\subsection{Final flux tables}

All the relatively (if the night was non-photometric but grey) and absolutely
(if the night was photometric) calibrated spectra will now have the correct
spectral shape. The absolutely calibrated spectra obtained in different nights
or with different telescopes for each star will be compared to study hidden
systematics (if any). Some of our targets belong to widely used
spectrophotometric datasets (see Section~\ref{sec-spss}), and will be our anchor
point to check our flux scale and to find potential problems. 

The relatively calibrated spectra will need a zeropoint correction. We will thus
use the version including telluric absorption features to derive synthetic B, V,
and R (and if available, I and U) magnitudes, and compare them with our calibrated
magnitudes (see previous section) to apply the necessary correction. Once this
procedure will be completed, all the spectra obtained for each SPSS will be
combined in one single spectrum: our final product. It will be necessary in many
cases to use synthetic spectra to calibrate the noisy edges, or the reddest
wavelength ranges, if they will not be properly cleaned from reddening. 

\begin{figure}
\includegraphics[width=\columnwidth]{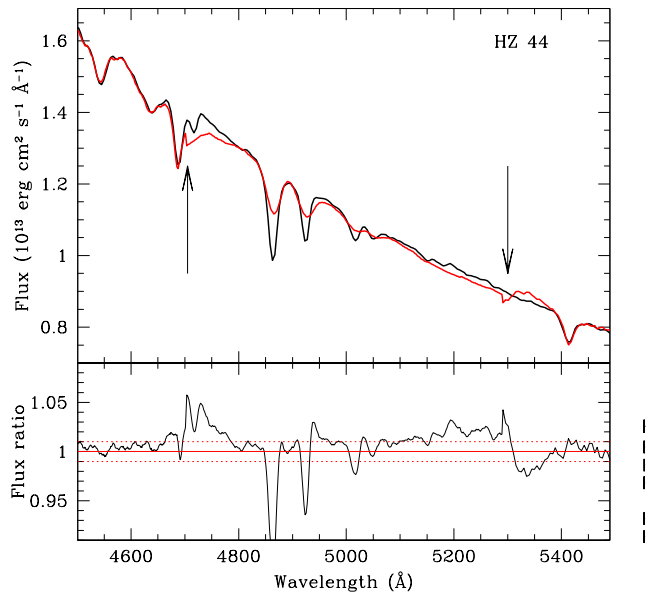}
\caption{Top panel: comparison of our preliminary spectrum of HZ~44 (thick black
line) with the CALSPEC tabulated spectrum (thick red line) in a region where we
found a discrepancy (marked by the two arrows), where small $\simeq$0.5--1.0\%
jumps in the CALSPEC spectrum are probably due to a mismatch of two different
spectra. Bottom panel: ratio between our spectrum and the CALSPEC spectrum;
perfect agreement (red line) and $\pm$1\% agreement (dotted red lines, our
requirement) are marked.}
\label{fig_hz44}
\end{figure}

As an example of the data quality, we show in Figure~\ref{fig_hz44} a test
performed to refine our reduction procedures, where a portion of the spectrum of
HZ44 observed in a photometric night is compared with the CALSPEC flux table. We
point out that this preliminary reduction did not include the proper extinction
curve, but a tabulated curve from \citet{sanchez07}; the telluric absorption
features were not removed \citep[we will use procedures similar to that
by][]{bessell99}; the red wavelengths are affected by fringing that we will beat
down by combining observations from different telescopes whenever possible; and
the extremes of the wavelength range are affected by poor S/N, so that we will
have to use synthetic spectra to calibrate those extremes. Even with these
limitations, we were able to meet the requirements (Section~\ref{sec-spss}),
because the residuals between our spectrum and the CALSPEC tabulated one were on
average lower than 1\%, with the exception of the low S/N red edge and of the
telluric absorption bands. However, some unsatisfactory jumps appeared in the
comparison, between 4000 and 6000~\AA, where our spectra have the highest S/N. As
shown in Figure~\ref{fig_hz44}  (top panel) and already noted by
\citet{bohlin01}, the jumps were due to a (minor) problem in the CALSPEC
spectrum, probably where two pieces of the spectrum were joined. 

Thus we were able to identify a defect in the CALSPEC spectrum of the order of
1-2\%, approximately, meeting the requirements (Section~\ref{sec-spss}). Similar
results were obtained on test reductions of other SPSS (observed with TNG, NTT,
and CAHA): GD71, GD153, and G191-B2B, our {\em Pillars}, which have the best
literature data available.

To produce our final flux tables, we will need to adjust model
spectra\footnote{We will make use of both atmosphere models from, e.g., the
MARCS, Kurucz, TLUSTY, and T\"ubingen sets
\citep{gustafsson08,castelli03,rauch03,lanz03,lanz07} or spectral libraries
\citep[e.g.,][]{sordo06,ringat12}.} to our observed spectra \citep[as done by,
e.g.,][]{bohlin07}. This technique has proven useful to identify and fix minor
problems on the spectra, and we will use it to correct for residuals from the
joining of different spectral pieces, sky subtraction, telluric features
correction, fringing, and imperfections at the spectral extremities, where the
S/N ratio is generally lower. Also, the use of models will allow us to
characterize our targets, thus providing spectral types, effective temperatures,
gravities, metallicities, and reddening. 

\begin{table*}
\begin{minipage}{175mm}
\caption{Notable rejected SPSS candidates\label{tab_rej}}
\begin{tabular}{@{\extracolsep{-10pt}}lccccll}
\hline
Star & RA (J2000) & Dec (J2000) & B & V & Type & Reason for rejection \\
     & (hh:mm:ss) & (dd:pp:ss) & (mag) & (mag) & \\
\hline
WD~0406+592 & 04:10:51.70\footnote{\citet{mccook99}; $^{b}$from 2MASS \citep{cutri03};
                         $^{c}$\citet{kharchenko01}; $^{d}$approximate spectral type 
                         from T$_{\rm eff}$ by \citet{carney94}; $^{e}$\citet{bohlin08};
                         $^{f}$UCAC3 \citep{zacharias09}; $^{g}$USNO-B catalogue \citep{monet03}.} 
                                &  +59:25:05.00$^{a}$ & 14.30$^{a}$ & 14.40$^{a}$ & DA$^{a}$  & Two close visual companions detected \\
G~192-41    & 06:44:26.34$^{b}$ &  +50:33:55.90$^{b}$ & 13.91$^{c}$ & 13.16$^{c}$ & G$^{d}$   & Suspected variable \\
WD~1148-230 & 11:50:38.80$^{a}$ & --23:20:34.00$^{a}$ & 11.49$^{a}$ & 11.76$^{a}$ & DA$^{a}$  & Two sets of coordinates and magnitudes in literature (see text) \\
1740346	    & 17:40:34.68$^{b}$ &  +65:27:14.80$^{b}$ & 12.68$^{e}$ & 12.48$^{e}$ & A5$^{e}$  & Variable, probably of $\delta$~Scuti type (CALSPEC standard) \\
WD~1911+135 & 19:13:38.68$^{b}$ &  +13:36:27.70$^{b}$ & 14.12$^{a}$ & 14.00$^{a}$ & DA3$^{a}$ & Crowded field \\
WD~1943+163 & 19:45:31.77$^{b}$ &  +16:27:39.60$^{b}$ & 13.96$^{a}$ & 13.99$^{a}$ & DA2$^{a}$ & Crowded field \\
WD~2046+396 & 20:48:08.18$^{f}$ &  +39:51:37.33$^{f}$ & 14.10$^{a}$ & 14.43$^{a}$ & DA1$^{a}$ & Crowded field  \\
WD~2058+181 & 21:01:16.49$^{b}$ &  +18 20 55.30$^{b}$ & 15.01$^{a}$ & 15.00$^{a}$ & DA4$^{a}$ & One close visual companion detected \\
WD~2256+313 & 22:58:39.44$^{b}$ &  +31:34:48.90$^{b}$ & 14.90$^{g}$ & 13.96$^{a}$ & --- & Fainter than expected \citep[see text,][]{oswalt88}\\
\hline
\end{tabular}
\end{minipage}
\end{table*}
\section{Preliminary results}
\label{sec-res}
\label{sec-rej}

We discuss in the following sections some preliminary results of our survey: a few
interesting cases of problematic candidates are described, and a list of notable
rejected SPSS candidates can be found in Table~\ref{tab_rej}; two stars showed
variability larger than $\pm$5~mmag in our short-term constancy monitoring.

\begin{figure}
\centering
\includegraphics[width=4.5cm]{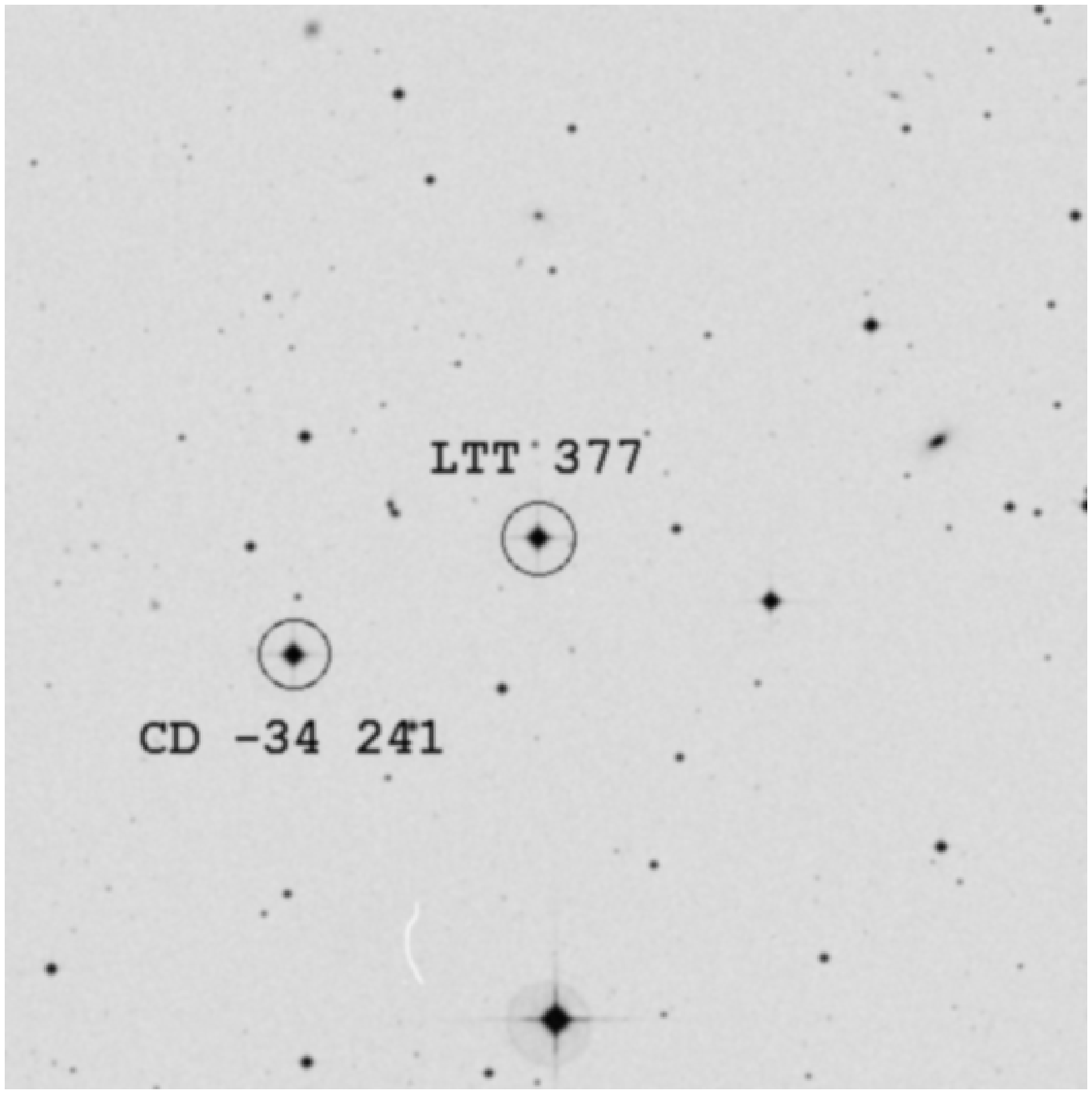}\includegraphics[width=4.5cm]{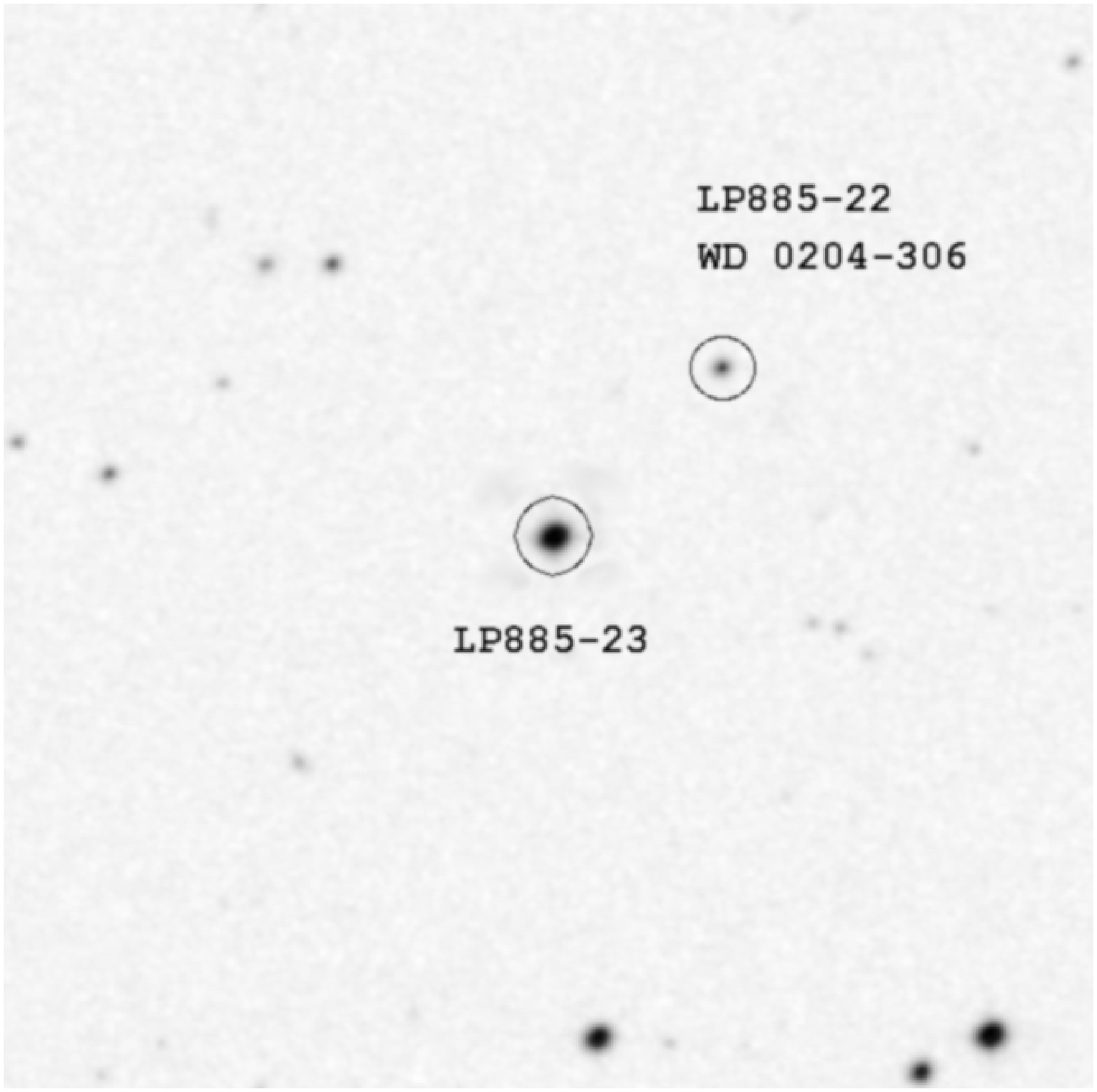}
\caption{Correct identifications of two candidate SPSS that were wrongly
identified in the literature. Left panel: the case of LTT~377, which was
confused with CD~-34~241; the image is 15' wide, North is up and East is left.
Right panel: the case of WD~0204-306, which was associated with LP~885-23
instead of LP~885-22; the image is 7' wide, North is up and East on the left.} 
\label{fig_ids1}
\end{figure}

\subsection{Identification and literature problems}
\label{sec-id}

Identification problems are common, especially when large databases are
automatically matched (as done within SIMBAD, for example), and when stars have
large proper motions. 

\begin{figure}
\includegraphics[width=4.5cm]{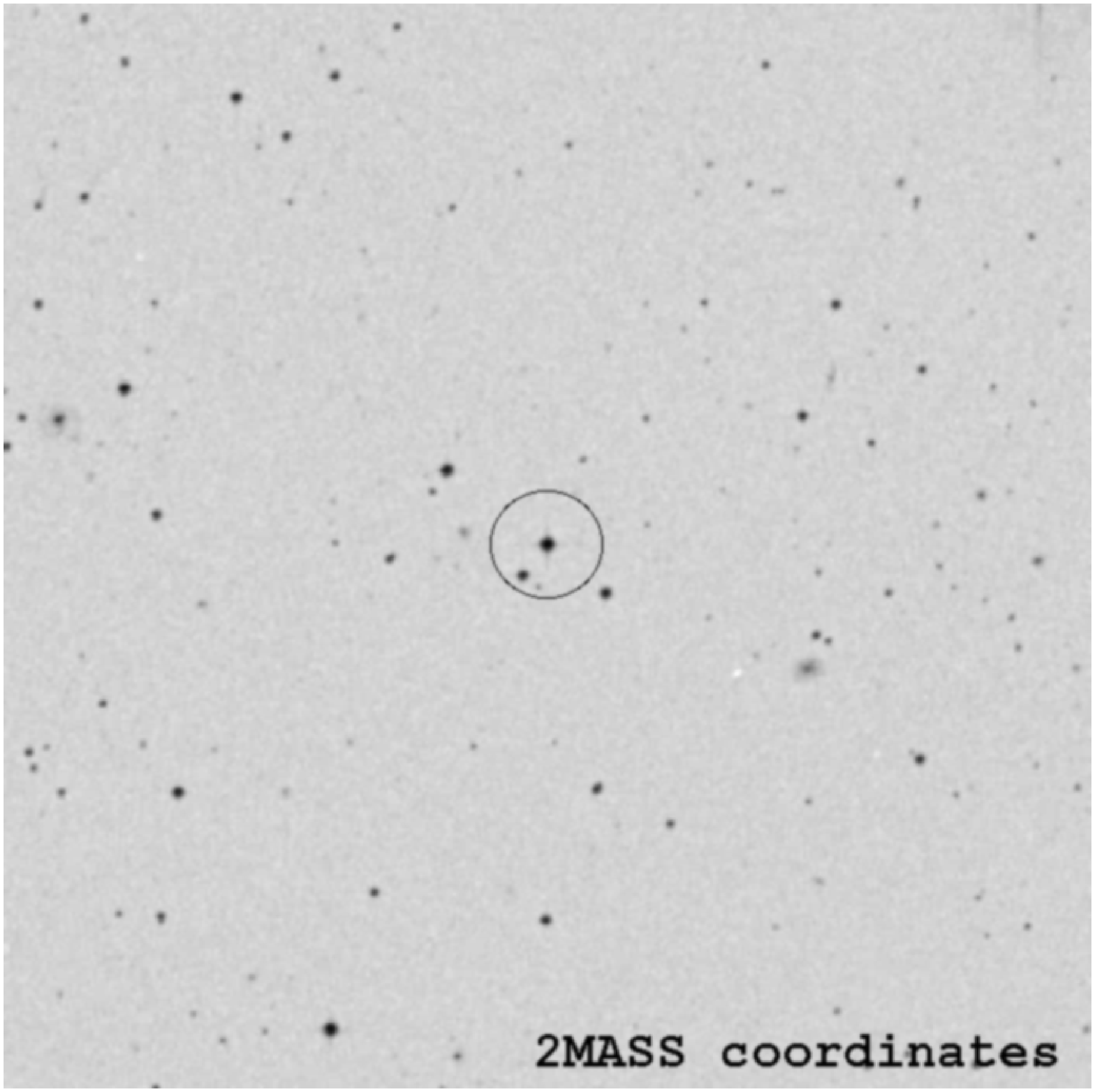}\includegraphics[width=4.5cm]{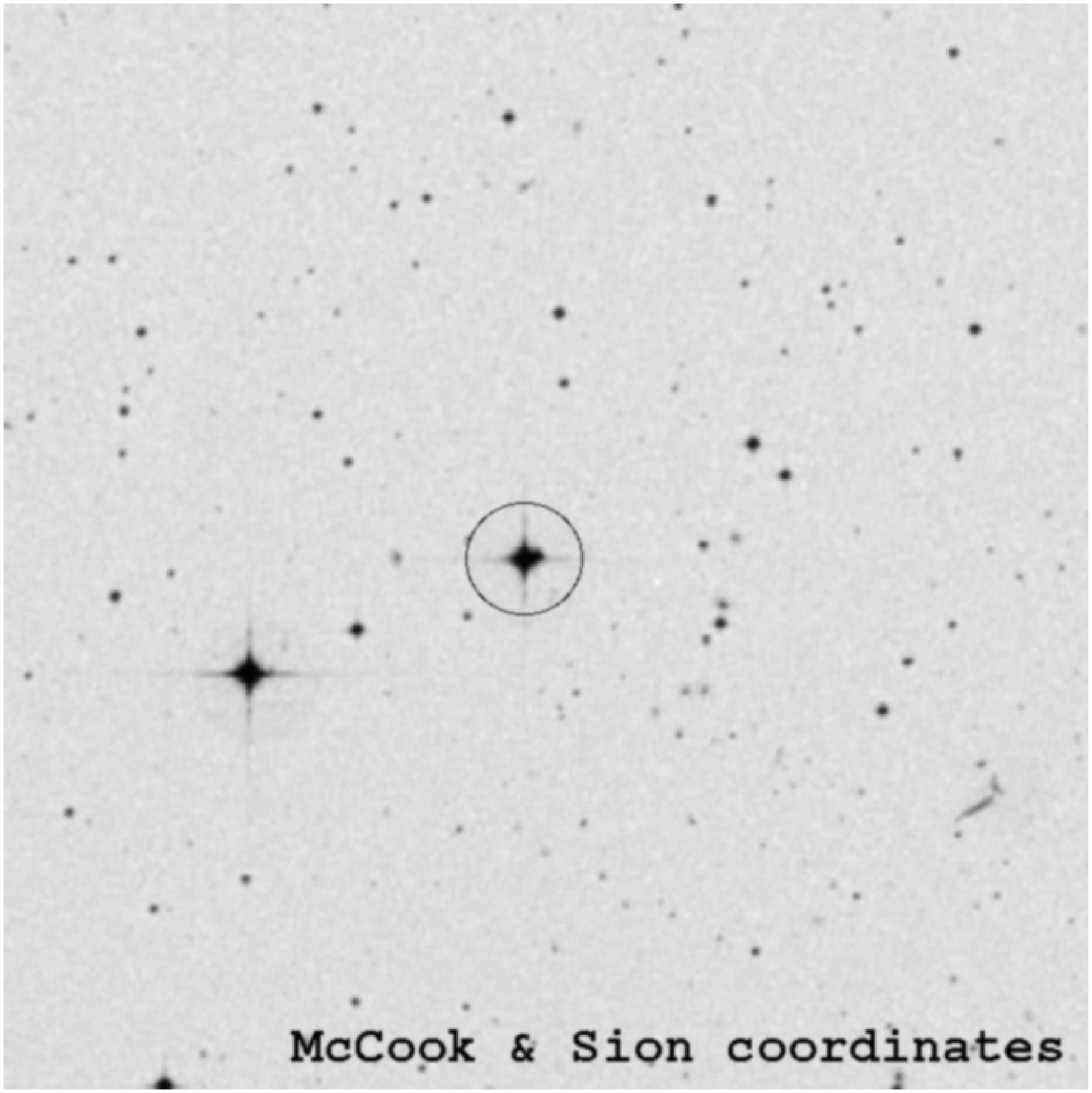}
\caption{The unsolved case of WD~1148-230. The finding chart on the left shows the
star that in SIMBAD is associated to WD~1148-230, at the coordinates reported by
2MASS \citep{cutri03}, the one on the right the star corresponding to the
WD~1148-230 coordinates by \citet{mccook99} and \citet{stys00}. Both images are
10' wide, North is up and East is left.}
\label{fig_wd1148}
\end{figure}

We found our first case when a discrepancy became evident between the
LTT~377\footnote{At the moment of writing, the SIMBAD database has been updated
and now the correct identification is reported.} literature spectrum
\citep{hamuy92,hamuy94} and our observed spectrum, which was more consistent with
an F type rather than the expected K spectral type. We contacted the SIMBAD and
ESO staff, because their sites reported the information from \citet{hamuy94} as
well, and we concluded that the ESO standard was not LTT~377, but another star
named CD~-34~241, of spectral type F. This was confirmed by older literature
papers like \citet{luyten57}, where LTT~377 was identified as CD~-34~239, and by
literature proper motions and coordinates. We could trace back the error to
\citet{stone83}, where the wrong association was probably done for the first
time, and then propagated down to SIMBAD and ESO. The correct identification of
both stars is shown in Figure~\ref{fig_ids1} (left panel)\footnote{The black and
white finding charts in Figures~\ref{fig_ids1} and \ref{fig_wd1148} were created
with the ESO SkyCat tool and images from the Digitized Sky Survey. SkyCat was
developed by ESO's Data Management and Very Large Telescope (VLT) Project
divisions with contributions from the Canadian Astronomical Data Center (CADC).}.
We decided to keep both stars in our candidates lists (see
Tables~\ref{tab_primaries} and \ref{tab_secondaries}).

A similar case was WD~0204-306 for which we obtained an unexpectedly red spectrum.
We traced literature identifications back to \citet{reid96}, who correctly
identified WD~0204-306 as associated with LP~885-23 (an M star) in a binary
system, with a separation of 73". At some point, the two stars got confused and in
SIMBAD WD~0204-306 (a white dwarf) was cross-identified with LP~885-23 (an M
star). Given the reported distance between the two stars, we identified
WD~0204-306 as LP~885-22, as shown in Figure~\ref{fig_ids1}. Also in this case,
having observations of both stars, we kept both in our {\em Secondary SPSS
candidates}. The mistake was reported to the SIMBAD staff and now the database is
corrected.

A more critical example was WD~1148-230 (Figure~\ref{fig_wd1148}), having very
different coordinates in the \citet{mccook99} catalogue \citep[coming from][and
reporting R.A.=11:50:38.8~h and Dec=--23:20:34~deg]{stys00} and in SIMBAD. The
SIMBAD coordinates were from the 2MASS catalogue \citep[][reporting
R.A.=11:50:06.09~h and Dec=--23:16:14.0~deg]{cutri03}. Magnitudes were also
significantly different. Unlike in the previous cases, we had insufficient
literature information to confirm one or the other identification, so we decided
to reject this SPSS candidate, although we suspect that the mistake resides in
the SIMBAD automatic association between WD~1148-230 by \citet{stys00} and the
2MASS catalogue.

Finally, we report on the case of WD~2256+313, which was reported to have
V=13.96~mag \citep{silvestri02,monet03}, but when observed in San Pedro M\'artir appeared
to be much fainter than that 
\citep[and of uncertain spectral type, see also][]{oswalt88}, possibly with V$>$15~mag, 
so was removed from our candidates list.

\subsection{Crowded fields and visual companions}

In a few cases candidates that appeared relatively isolated on the available
finding charts turned out to be in a crowded area where no aperture photometry or
reliable wide slit spectroscopy could be performed from the ground, or showed
previously unseen companions. Generally, stars with high proper motion could
appear isolated in some past finding chart, but later moved too close to another
star to be safely observed from the ground. 

\begin{figure}
\includegraphics[width=4cm,height=4cm]{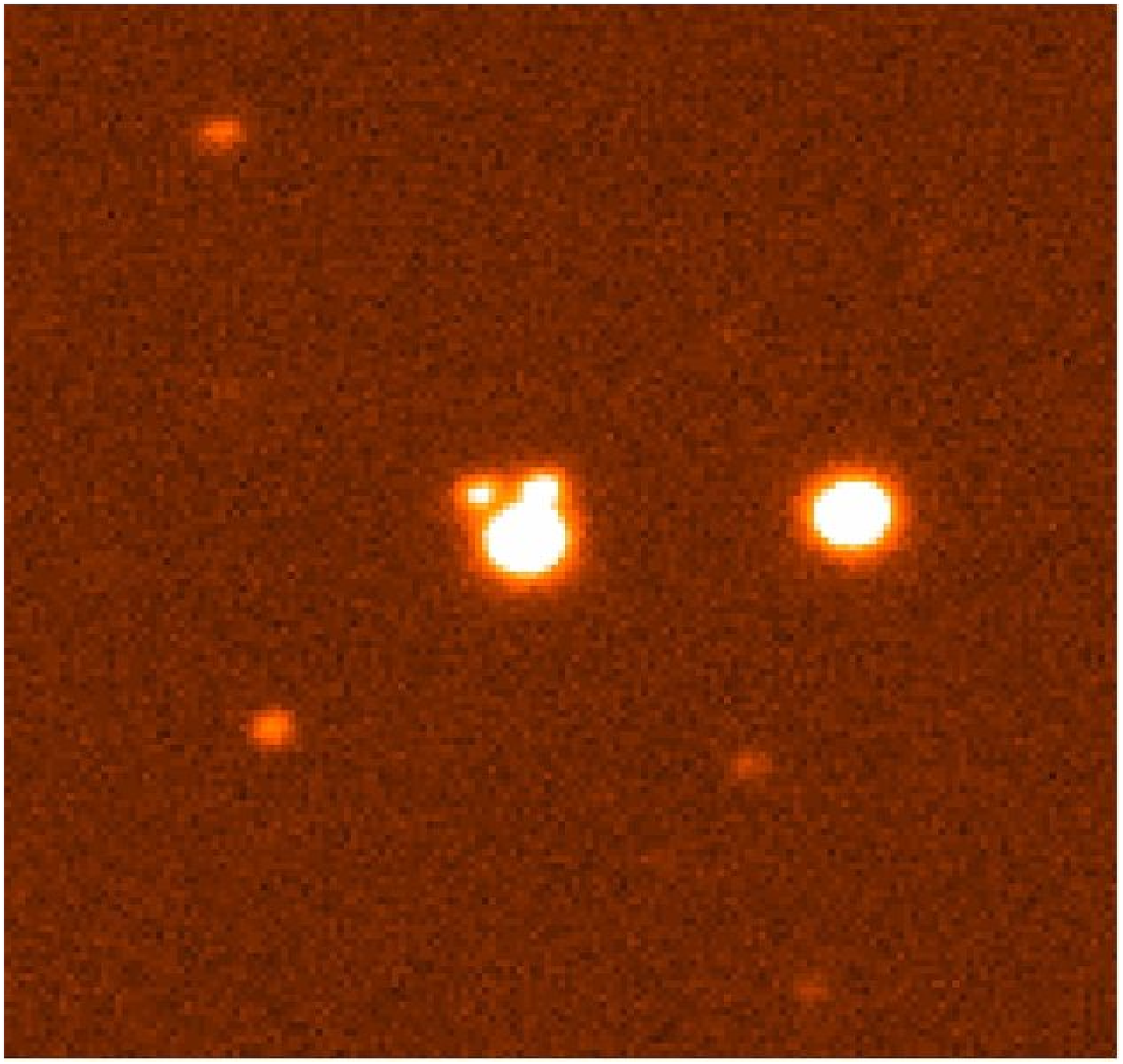}\hspace{0.5cm}\includegraphics[width=4cm,height=4cm]{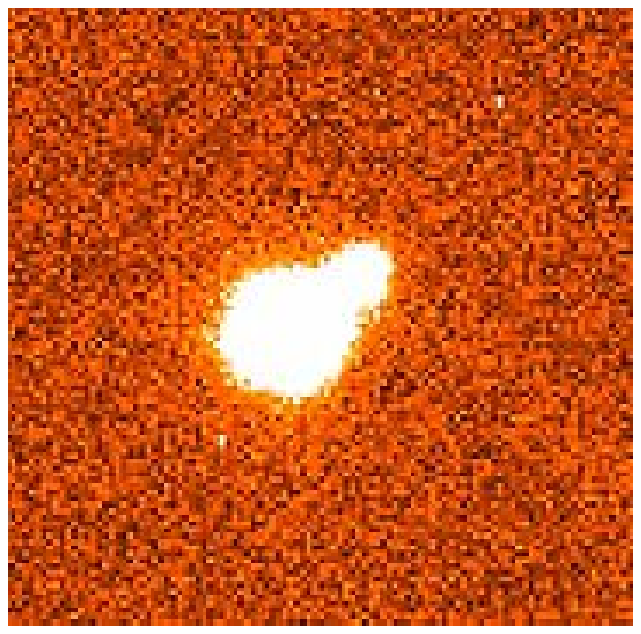}
\caption{Image cutout of candidate WD~0406+592 (left panel) observed with
DOLoRes@TNG in the R band, showing two close companions; similarly, a cutout of
candidate WD~2058+181 (right panel), observed in San Pedro M\'artir in the R
band, shows a close companion.}
\label{fig_crowd}
\end{figure}

One example of candidate which appeared relatively isolated judging from the 
\citet{mccook99} finding charts, but turned out to be in a crowded field when 
observed at San Pedro M\'artir was WD~1911+135, that was promptly rejected,
together with WD~1943+163 and WD~2046+396. Examples of candidates showing the
presence of previously unknown and relatively bright companions were WD~0406+592
and WD~2058+181 (Figure~\ref{fig_crowd}). These stars do not have a particularly
high proper motion, and appeared easy to identify on the corresponding finding
charts, so we did not expect them to show close visual companions, when observed
from the TNG and San Pedro M\'artir, respectively. Both stars were rejected.

\subsection{Variability} 
\label{sec-var}

\begin{figure}
\includegraphics[angle=270, width=\columnwidth]{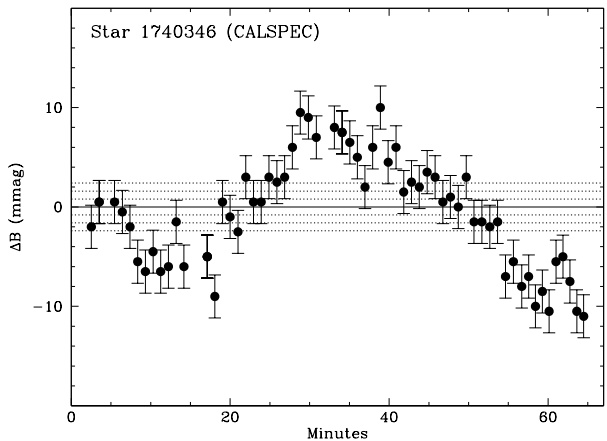}
\caption{Our best lightcurve for the CALSPEC standard 1740346 (obtained with
BFOSC in Loiano on 1 September 2010), originally one of our {\em Primary SPSS}
candidates. The average of all field-stars magnitude differences (i.e., zero) is
marked with a solid line, while the $\pm$1, 2, and 3~$\sigma$ variations are
marked with dotted lines.}
\label{fig_var}
\end{figure}

Our auxiliary campaign started giving results as far as the short-term constancy
monitoring (1--2~h) is concerned. The ability of one lightcurve to detect
magnitude variations is measured using the spread of reference star's magnitude
differences. These appear as 1, 2, and 3~$\sigma$ limits in Figure~\ref{fig_var},
where we present the differential lightcurves our only {\em confirmed} variable
star. 

Star 1740346, one of the currently used CALSPEC standards and one of our {\em
Primary SPSS} candidates, showed variability with an amplitude of
10~$\pm$~0.8~mmag in B band when observed with BFOSC@Cassini in Loiano, on 1
September 2010; with DOLoRes@TNG, on 31 September 2009; and with BFOSC@Cassini,
on 26 May 2009. The variability period is 50~min, approximately, thus showing
properties typical of $\delta$~Scuti variables. A preliminary determination of
1740346 parameters can be found in \citet{marinoni11}, using literature data and
stellar models, resulting in a mass of $\simeq$1.3~M$_{\odot}$, an effective
temperature of $\simeq$8300~K, and a distance of $\simeq$750~pc. These parameters
are also compatible with a $\delta$~Scuti type star. We are gathering detailed
follow-up observations and a complete characterization of star 1740346 will be
the subject of a forthcoming paper (Marinoni et al., in preparation). The
differential lightcurve is presented in Figure~\ref{fig_var} (top panel).

\section{Summary and conclusions}

We have described a large (more than 400 nights) ground-based survey which started
in 2007 and is expected to end in 2013--2014, aimed at building a grid of SPSS for
the flux calibration of Gaia spectra and magnitudes. The technical complexity of
Gaia requires a large ($\simeq$200) set of SPSS flux tables, calibrated in flux
with high precision ($\simeq$1\%) and accuracy ($\simeq$3\% with respect to Vega),
and covering a range of spectral types. SPSS candidates need to be monitored for
constancy (within $\pm$5~mmag) to ensure the quoted precision in the final
calibration. 

We discussed the adopted calibration strategy, the selection requirements and a
list of candidate SPSS. A brief overview of the adopted data reduction and
analysis procedures was also presented, and more details will be discussed in a
series of future papers dealing with all technical aspects, data products,
photometric catalogues, flux tables, and lightcurves. Some preliminary results
were presented, showing the data quality, a few problematic cases of candidate
SPSS that were rejected because of identification problems, close companions, and
variability. In particular, we detected a new variable star, a CALSPEC standard
which is most probably a $\delta$ Scuti variable; follow-up observations for its
characterization are ongoing.

All data products will be eventually made public together with each Gaia data
release, within the framework of the DPAC (Data Processing and Analysis
Consortium) publication policies. At the moment the accumulated data and
literature information are stored locally and can be accessed upon request.

\section*{Acknowledgments}

We would like to acknowledge the support of the INAF (Istituto Nazionale di
Astrofisica) and specifically of the Bologna Observatory; of the ASI (Agenzia
Spaziale Italiana) under contracts to INAF I/037/08/0 and I/058/10/0, dedicated
to the Gaia mission, and the Italian participation to DPAC (Data Analysis and
Processing Consortium). This work was supported by the /MICINN/ (Spanish
Ministry of Science and Innovation) --- FEDER through grant AYA2009-14648-C02-01
and CONSOLIDER CSD2007-00050. EP acknowledges the hospitality os the ASDC (ASI
Sciece Data Center), where part of this work was carried out. We warmly thank
the technical staff of the San Pedro M\'artir, Calar Alto, Loiano, La Silla NTT
and REM, and Roque de Los Muchachos TNG observatories.

We made use of the following softwares and online databases (in alphabetical
order): 2MASS, CALSPEC, CataXcorr, ESO-DSS, ESO Skycat tool, IRAF, Kawka webpage,
MILES, NGSL, SAOImage DS9, SDSS and SEGUE, SExtractor, SIMBAD, SuperMongo, UCAC3,
USNO catalogues, Villanova White Dwarf Catalogue. We thank G.~S.~Aldering,
M.~.A.~ Barstow, M.~E.~Kayser, and A.~Korn for sharing their information with us.
We also thank M.~Bessell, who was the referee of this paper and provided
extremely useful comments not only to improve the paper, but for the whole
project.

The survey presented in this paper relies on data obtained at ESO (proposals
182.D-0287, 086.D-0176, 087.D-0213, and 089.D-0077), Calar Alto (proposals
H07-2.2-024, F08-2.2-043, H08-2.2-041, F10-2.2-027, H10-2.2-042, H10-2.2-042, and
F12-2.2-034), TNG (proposals AOT16\_37, AOT17\_3, AOT18\_14, AOT19\_14, AOT20\_41,
and AOT21\_1), Loiano (10 accepted proposals starting from June 2007), San Pedro 
M\'artir (7 accepted proposals starting from October 2007), and REM (proposals 
AOT16\_16012, AOT17\_17012, AOT18\_18002, AOT19\_19010, AOT20\_78, AOT21\_2,
AOT22\_18, AOT23\_7, AOT24\_21).


\newpage
\onecolumn
\renewcommand{\thetable}{\arabic{table}}
\setcounter{footnote}{0}

\begin{longtable}{@{\extracolsep{-10pt}}lccccllccccl}
\caption{\label{tab_secondaries} Secondary SPSS candidates}\\
\hline
Star & RA (J2000) & Dec (J2000)	& B & V & Type & Star & RA (J2000) & Dec (J2000) & B & V & Type \\
     & (hh:mm:ss) & (dd:pp:ss) & (mag) & (mag) &      & & (hh:mm:ss) & (dd:pp:ss) & (mag) & (mag) & \\
\hline
\endfirsthead
\caption{continued.}\\
\hline
Star & RA (J2000) & Dec (J2000)	& B & V & Type & Star & RA (J2000) & Dec (J2000) & B & V & Type \\
     & (hh:mm:ss) & (dd:pp:ss) & (mag) & (mag) &      & & (hh:mm:ss) & (dd:pp:ss) & (mag) & (mag) & \\
\hline
\endhead
\hline
\endfoot  
WD~2359-434   & 00:02:10.77\footnote{2MASS survey \citep{cutri03}; $^{2}$\citet{mccook99} compilation and online updates;
                            $^{3}$SDSS seventh data release \citep{SDSS7}; $^{4}$SDSS, derived with the SEGUE pipeline 
                            \citep{lee08} and the transformations by \citep{lupton05}; $^{5}$SDSS, derived with the SEGUE 
                            pipeline \citep{lee08}; $^{6}$\citet{vanleeuwen07}; $^{7}$\citet{koen10}; $^{8}$\citet{gray06};
                            $^{9}$Tycho-2 catalogue of bright sources \citep{hog00}; $^{10}$\citet{carney87};
                            $^{11}$from T$_{\rm{eff}}$ by \citet{laird88}; $^{12}$\citet{jenkins09}; $^{13}$\citet{garces11};
                            $^{14}$\citet{hog98}, for approximate Johnson magnitudes the formulae V=VT--0.090*(BT--VT) and 
                            B--V=0.850*(BT--VT) where used; $^{15}$\citet{kharchenko01}; $^{16}$\citet{bidelman85}; 
                            $^{17}$\citet{salim03}; $^{18}$``Subdwarf database" \citep{ostensen06}; $^{19}$misclassified as a 
                            WD by \citet{mccook99} according to \citet{stroeer07}; $^{20}$from T$_{\rm{eff}}$ by 
                            \citet{carney94}; $^{21}$\citet{galadi00}; $^{22}$Henry Draper Catalogue \citep{cannon93};
                            $^{23}$\citet{caballero07}; $^{24}$\citet{carney94}; $^{25}$UCAC3 \citep{zacharias09};
                            $^{26}$Hipparcos input catalogue \citep{turon93}; $^{27}$\citet{marshall07}; 
                            $^{28}$\citet{cenarro07}; $^{29}$\citet{lepine05}; $^{30}$from T$_{\rm{eff}}$ by 
                            \citet{latham02}; $^{31}$\citet{bicay00}; $^{32}$\citet{landolt92}; $^{33}$\citet{zickgraf03};
                            $^{34}$\citet{landolt07}; $^{35}$\citet{giclas71}; $^{36}$\citet{ivanov08};
                            $^{37}$\citet{mermilliod94}; $^{38}$\citet{buscombe95}; $^{39}$\citet{drilling79};
                            $^{40}$\citet{pesch76}; $^{41}$\citet{vanaltena95}; $^{42}$\citet{kharchenko09};
                            $^{43}$7th SDSS photometric data release \citep{adelman09}; $^{44}$\citet{tanabe08};
                            $^{45}$\citet{downes01}; $^{46}$\citet{monet03}; $^{47}$\citet{zapatero04};
                            $^{48}$\citet{greenstein84}; $^{49}$\citet{zacharias05}; $^{50}$\citet{malina94};
                            $^{51}$\citet{fleming96}; $^{52}$\citet{koester01}; $^{53}$\citet{roeser88};
                            $^{54}$\citet{wegner73}; $^{55}$\citet{vennes97}; $^{56}$\citet{lee84};
                            $^{57}$Stetson standard in M~5 \citep{stetson00}; data available at 
                            http://cadcwww.dao.nrc.ca/community/STETSON/standards; $^{58}$\citet{endl06};
                            $^{*}$possible identification problem, see also Section~\ref{sec-id}.}   
                                    & --43:09:56.02$^{1}$   & 13.12$^{2}$   & 13.05$^{2}$   & DA5$^{2}$   & HD~271759     & 06:00:41.34$^{14}$  & --66:03:14.03$^{14}$  & 11.00$^{9}$   & 11.20$^{9}$   & A2$^{22}$   \\  
WD~0004+330   & 00:07:32.26$^{1}$   &  +33:17:27.60$^{1}$   & 13.57$^{2}$   & 13.85$^{2}$   & DA1$^{2}$   & HD~271783     & 06:02:11.36$^{9}$   & --66:34:59.13$^{9}$   & 12.63$^{9}$   & 12.23$^{9}$   & F5$^{22}$   \\  
SDSS~03932    & 00:07:52.22$^{3}$   &  +14:30:24.72$^{3}$   & 15.37$^{4}$   & 15.07$^{4}$   & A0$^{5}$    & HIP~28618     & 06:02:27.88$^{6}$   & --66:47:28.68$^{6}$   & 12.20$^{9}$   & 12.30$^{9}$   & B8$^{22}$   \\  
WD~0009+501   & 00:12:14.80$^{1}$   &  +50:25:21.40$^{1}$   & 14.78$^{2}$   & 14.36$^{2}$   & DA8$^{2}$   & WD0604-203    & 06:06:13.39$^{1}$   & --20:21:07.20$^{1}$   & 11.75$^{23}$  & 11.80$^{23}$  & DA$^{23}$   \\  
WD~0018-267   & 00:21:30.73$^{1}$   & --26:26:11.46$^{1}$   &  ---          & 13.80$^{2}$   & DA9$^{2}$   & WD0621-376    & 06:23:12.63$^{1}$   & --37:41:28.01$^{1}$   & 11.76$^{2}$   & 12.09$^{2}$   & DA1$^{2}$   \\  
SDSS~03532    & 00:24:38.62$^{3}$   & --01:11:39.75$^{3}$   & 15.14$^{4}$   & 15.04$^{4}$   & A0$^{5}$    & WD0644+375    & 06:47:37.99$^{6}$   &  +37:30:57.07$^{6}$   & 11.99$^{2}$   & 12.08$^{2}$   & DA2$^{2}$   \\  
WD~0038+555   & 00:41:21.99$^{1}$   &  +55:50:08.40$^{1}$   & 14.10$^{2}$   & 14.08$^{2}$   & DQ5$^{2}$   & WD0646-253    & 06:48:56.09$^{1}$   & --25:23:47.00$^{1}$   & 13.30$^{2}$   & 13.40$^{2}$   & DA2$^{2}$   \\  
LTT~377       & 00:41:30.47$^{6}$   & --33:37:32.03$^{6}$   & 11.97$^{7}$   & 10.53$^{7}$   & K9$^{8}$    & G193-26       & 07:03:26.29$^{1}$   &  +54:52:06.00$^{1}$   & 13.59$^{24}$  & 13.02$^{24}$  & G$^{20}$    \\  
WD~0046+051   & 00:49:09.90$^{6}$   &  +05:23:19.01$^{6}$   & 12.93$^{2}$   & 12.39$^{2}$   & DZ7$^{2}$   & WD0713+584    & 07:17:36.26$^{6}$   &  +58:24:20.51$^{6}$   & 12.06$^{9}$   & 12.02$^{9}$   & DA4$^{2}$   \\  
WD~0047-524   & 00:50:03.68$^{1}$   & --52:08:15.60$^{1}$   & 14.19$^{2}$   & 14.20$^{2}$   & DA2$^{2}$   & WD0721-276    & 07:23:20.10$^{1}$   & --27:47:21.60$^{1}$   & 13.50$^{2}$   & 13.40$^{2}$   & DA1$^{2}$   \\  
WD~0050-332   & 00:53:17.44$^{1}$   & --32:59:56.60$^{1}$   & 13.11$^{2}$   & 13.36$^{2}$   & DA1$^{2}$   & WD0749-383    & 07:51:32.58$^{25}$  & --38:28:36.41$^{25}$  & 13.53$^{2}$   & 13.66$^{2}$   & DA$^{2}$    \\  
WD~0104-331   & 01:06:46.86$^{1}$   & --32:53:12.45$^{1}$   & 13.28$^{2}$   & 13.57$^{2}$   & DAZ3$^{2}$  & G251-54       & 08:11:06.24$^{6}$   &  +79:54:29.57$^{6}$   & 10.58$^{26}$  & 10.01$^{26}$  & G0$^{26}$   \\  
WD~0106-358   & 01:08:20.80$^{2}$   & --35:34:43.00$^{2}$   & 14.54$^{2}$   & 14.72$^{2}$   & DA2$^{2}$   & GJ2066        & 08:16:07.98$^{6}$   &  +01:18:09.26$^{6}$   & 11.63$^{7}$   & 10.09$^{7}$   & M2$^{12}$   \\  
WD~0109-264   & 01:12:11.65$^{9}$   & --26:13:27.69$^{9}$   & 12.91$^{2}$   & 13.15$^{2}$   & DA1$^{2}$   & G114-25       & 08:59:03.37$^{6}$   & --06:23:46.19$^{6}$   & 12.52$^{27}$  & 11.97$^{27}$  & F7$^{28}$   \\  
WD~0123-262   & 01:25:24.45$^{1}$   & --26:00:43.90$^{1}$   & 15.35$^{2}$   & 14.95$^{2}$   & DC$^{2}$    & WD0859-039    & 09:02:17.30$^{1}$   & --04:06:55.45$^{1}$   & 13.02$^{2}$   & 13.19$^{2}$   & DA2$^{2}$   \\  
G245-31       & 01:38:39.39$^{1}$   &  +69:38:01.50$^{1}$   & 15.26$^{10}$  & 14.50$^{10}$  & K$^{11}$    & WD0912+536    & 09:15:56.23$^{1}$   &  +53:25:24.90$^{1}$   & 14.19$^{2}$   & 13.85$^{2}$   & DB/DC$^{2}$ \\  
WD~0134+833   & 01:41:28.74$^{1}$   &  +83:34:58.90$^{1}$   & 12.88$^{2}$   & 13.11$^{2}$   & DA2$^{2}$   & WD0943+441    & 09:46:39.08$^{1}$   &  +43:54:52.37$^{1}$   & 13.19$^{2}$   & 13.12$^{2}$   & DA4$^{2}$   \\  
GJ70 	      & 01:43:20.18$^{6}$   &  +04:19:17.97$^{6}$   & 12.45$^{7}$   & 10.92$^{7}$   & M2$^{12}$   & G43-5         & 09:49:51.59$^{9}$   &  +06:36:35.64$^{9}$   & 12.90$^{29}$  & 12.48$^{29}$  & K$^{30}$    \\  
G72-34 	      & 01:46:03.66$^{1}$   &  +35:54:49.40$^{1}$   & 13.84$^{10}$  & 12.98$^{10}$  & K$^{11}$    & WD0954-710    & 09:55:22.89$^{1}$   & --71:18:08.31$^{1}$   & 13.60$^{2}$   & 13.48$^{2}$   & DA4$^{2}$   \\  
WD~0147+674   & 01:51:10.29$^{1}$   &  +67:39:31.30$^{1}$   & 14.17$^{2}$   & 14.42$^{2}$   & DA2$^{2}$   & G236-30       & 10:28:48.37$^{1}$   &  +62:59:45.00$^{1}$   & 13.62$^{15}$  & 12.87$^{15}$  & G5$^{15}$   \\  
WD~0148+467   & 01:52:02.96$^{6}$   &  +47:00:06.65$^{6}$   & 12.50$^{2}$   & 12.44$^{2}$   & DA3$^{2}$   & WD1029+537    & 10:32:10.26$^{31}$  &  +53:29:36.40$^{31}$  & 14.18$^{2}$   & 14.46$^{2}$   & DA1$^{2}$   \\  
WD~0204-306$^{*}$& 02:07:02.28$^{1}$& --30:23:32.20$^{1}$   & ---           & 16.18$^{13}$  & DA$^{2}$    & WD1031-114    & 10:33:42.76$^{25}$  & --11:41:38.35$^{25}$  & 12.85$^{2}$   & 13.03$^{2}$   & DA2$^{2}$   \\  
LP885-23$^{*}$& 02:07:06.33$^{1}$   & --30:24:22.90$^{1}$   & ---           & 13.06$^{13}$  & M0$^{13}$   & WD1034+001    & 10:37:03.81$^{1}$   & --00:08:19.30$^{1}$   & 12.86$^{32}$  & 13.23$^{32}$  & DOZ1$^{2}$  \\  
WD~0214+568   & 02:17:33.52$^{1}$   &  +57:06:47.50$^{1}$   & 13.56$^{2}$   & 13.68$^{2}$   & DA2$^{2}$   & WD1041+580    & 10:44:46.10$^{33}$  &  +57:44:35.00$^{33}$  & 14.37$^{2}$   & 14.60$^{2}$   & DA1$^{2}$   \\  
WD~0227+050   & 02:30:16.62$^{6}$   &  +05:15:50.68$^{6}$   & 12.75$^{7}$   & 12.80$^{7}$   & DA3$^{2}$   & WD1053-550    & 10:55:13.54$^{1}$   & --55:19:05.20$^{1}$   & 14.42$^{2}$   & 14.32$^{2}$   & DA4$^{2}$   \\  
WD~0302+621   & 03:06:16.69$^{1}$   &  +62:22:22.68$^{1}$   & 15.17$^{2}$   & 14.95$^{2}$   & DA4/6$^{2}$ & WD1056-384    & 10:58:20.11$^{1}$   & --38:44:25.10$^{1}$   & 13.86$^{34}$  & 14.05$^{34}$  & DA2$^{2}$   \\  
WD~0316-849   & 03:09:59.89$^{14}$  & --84:43:21.14$^{14}$  & 11.62$^{14}$  & 10.55$^{14}$  & DAH$^{2}$   & G146-76       & 10:59:57.48$^{9}$   &  +44:46:43.75$^{9}$   & 11.15$^{9}$   & 10.47$^{9}$   & G/K $^{20}$ \\  
G174-44       & 03:17:23.31$^{1}$   &  +52:17:42.40$^{1}$   & 14.49$^{15}$  & 13.75$^{15}$  & K0$^{16}$   & WD1104+602    & 11:07:42.80$^{1}$   &  +59:58:29.90$^{1}$   & 13.78$^{2}$   & 13.80$^{2}$   & DA3$^{2}$   \\  
HG7-15 	      & 03:48:11.86$^{6}$   &  +07:08:46.47$^{6}$   & 12.11$^{29}$  & 10.65$^{29}$  & M1$^{58}$   & WD1105-048    & 11:07:59.95$^{1}$   & --05:09:25.90$^{1}$   & 13.09$^{32}$  & 13.06$^{32}$  & DA3$^{2}$   \\
WD~0435-088   & 04:37:47.42$^{17}$  & --08:49:10.70$^{17}$  & 14.10$^{2}$   & 13.77$^{2}$   & DQ7$^{2}$   & G10-4         & 11:10:60.00$^{6}$   &  +06:25:11.51$^{6}$   & 12.13$^{35}$  & 11.41$^{35}$  & K$^{20}$    \\  
WD~0446-789   & 04:43:46.47$^{1}$   & --78:51:50.40$^{1}$   & 13.36$^{2}$   & 13.47$^{2}$   & DA3$^{2}$   & G254-24       & 11:32:23.31$^{6}$   &  +76:39:18.03$^{6}$   & 12.18$^{36}$  & 11.53$^{36}$  & G0$^{16}$   \\  
WD~0447+176   & 04:50:13.52$^{6}$   &  +17:42:06.21$^{6}$   & 12.63$^{17}$  & 12.65$^{18}$  & sdO$^{19}$  & WD1134+300    & 11:37:05.10$^{6}$   &  +29:47:58.34$^{6}$   & 12.41$^{34}$  & 12.49$^{34}$  & DA2$^{2}$   \\  
WD~0455-282   & 04:57:13.90$^{2}$   & --28:07:54.00$^{2}$   & 13.63$^{2}$   & 13.95$^{2}$   & DA1$^{2}$   & SDSS~09310    & 11:38:02.62$^{3}$   &  +57:29:23.89$^{3}$   & 15.24$^{4}$   & 14.99$^{4}$   & A0/F3$^{5}$ \\  
WD~0501-289   & 05:03:55.51$^{2}$   & --28:54:34.57$^{2}$   & 13.55$^{2}$   & 13.90$^{2}$   & DO$^{2}$    & G10-54        & 11:49:48.20$^{1}$   &  +06:08:52.14$^{1}$   & 13.17$^{37}$  & 12.57$^{37}$  & G$^{20}$    \\  
G191-52       & 05:44:43.55$^{1}$   &  +56:15:30.80$^{1}$   & 14.02$^{15}$  & 13.26$^{15}$  & G$^{20}$    & WD1153-484    & 11:56:11.43$^{1}$   & --48:40:03.18$^{1}$   & 12.65$^{2}$   & 12.85$^{2}$   & DA2$^{2}$   \\  
U1050-027792  & 05:52:18.18$^{1}$   &  +15:51:52.70$^{1}$   & 14.42$^{21}$  & 13.70$^{21}$  & ---         & WD1210+533    & 12:13:24.64$^{1}$   &  +53:03:57.36$^{1}$   & 13.78$^{2}$   & 14.12$^{2}$   & DAO1$^{2}$  \\  
WD~0552-041   & 05:55:09.53$^{17}$  & --04:10:07.10$^{17}$  & 15.50$^{2}$   & 14.45$^{2}$   & DC/DZ$^{2}$ & WD1211-169    & 12:14:10.53$^{14}$  & --17:14:20.19$^{14}$  & 11.04$^{15}$  & 10.13$^{15}$  & DAH$^{2}$   \\  
HD~270422     & 05:56:47.74$^{14}$  & --66:39:05.27$^{14}$  & 10.92$^{9}$   & 10.05$^{9}$   & G0$^{22}$   & GJ459.3       & 12:19:24.09$^{6}$   &  +28:22:56.52$^{6}$   & 12.06$^{26}$  & 10.62$^{26}$  & M2$^{26}$   \\  
HD~270477     & 05:59:33.36$^{14}$  & --67:01:13.72$^{14}$  & 10.73$^{9}$   & 10.28$^{9}$   & F8$^{22}$   & SDSS~12720    & 12:22:41.66$^{3}$   &  +42:24:43.66$^{3}$   & 15.18$^{4}$   & 15.04$^{4}$   & A0/F2$^{5}$ \\  
HD~271747     & 05:59:58.62$^{9}$   & --66:06:08.91$^{9}$   & 11.82$^{9}$   & 11.29$^{9}$   & G0$^{22}$   & WD1223-659    & 12:26:42.02$^{1}$   & --66:12:18.70$^{1}$   & 14.37$^{2}$   & 13.97$^{2}$   & DA7$^{2}$   \\  
WD1234+481    & 12:36:45.18$^{1}$   &  +47:55:22.34$^{1}$   & 14.09$^{2}$   & 14.42$^{2}$   & DA1$^{2}$   & WD2047+372    & 20:49:06.69$^{1}$   &  +37:28:13.90$^{1}$   & 13.07$^{2}$   & 12.93$^{2}$   & DA3$^{2}$   \\
SA 104-428    & 12:41:41.28$^{1}$   & --00:26:26.20$^{1}$   & 13.58$^{32}$  & 12.63$^{32}$  & G8$^{38}$   & WD2111+498    & 21:12:44.05$^{1}$   &  +50:06:17.80$^{1}$   & 12.84$^{2}$   & 13.08$^{2}$   & DA1$^{2}$   \\
SA 104-490    & 12:44:33.46$^{1}$   & --00:25:51.70$^{1}$   & 13.07$^{32}$  & 12.57$^{32}$  & G3$^{39}$   & WD2105-820    & 21:13:13.90$^{2}$   & --81:49:04.00$^{2}$   & 13.82$^{2}$   & 13.61$^{2}$   & DA5$^{2}$   \\
G14-24        & 13:02:01.58$^{1}$   & --02:05:21.42$^{1}$   & 13.52$^{27}$  & 12.81$^{27}$  & K0$^{20}$   & WD2111+261    & 21:13:45.93$^{1}$   &  +26:21:33.20$^{1}$   & 14.92$^{2}$   & 14.68$^{2}$   & DA6$^{2}$   \\
GJ2097        & 13:07:04.31$^{25}$  &  +20:48:38.54$^{25}$  & 14.10$^{40}$  & 12.54$^{40}$  & M1$^{12}$   & WD2117+539    & 21:18:56.27$^{9}$   &  +54:12:41.25$^{9}$   & 12.40$^{2}$   & 12.33$^{2}$   & DA3$^{2}$   \\
SDSS~08393    & 13:10:32.07$^{3}$   &  +54:18:33.66$^{3}$   & 15.30$^{4}$   & 15.08$^{4}$   & A0/F3$^{5}$ & WD2115-560    & 21:19:36.52$^{1}$   & --55:50:14.20$^{1}$   & 14.43$^{2}$   & 14.28$^{2}$   & DAZ5$^{2}$  \\
GJ507.1       & 13:19:40.13$^{6}$   &  +33:20:47.49$^{6}$   & 12.10$^{29}$  & 10.57$^{29}$  & M2$^{12}$   & WD2122+282    & 21:24:58.30$^{2}$   &  +28:26:05.00$^{2}$   & 13.80$^{2}$   & 14.00$^{2}$   & DA0$^{55}$  \\
WD1319+466    & 13:21:15.08$^{1}$   &  +46:23:23.68$^{1}$   & 14.55$^{2}$   & 14.55$^{2}$   & DA3$^{2}$   & WD2136+828    & 21:33:43.25$^{1}$   &  +83:03:32.40$^{1}$   & 13.01$^{2}$   & 13.02$^{1}$   & DA3$^{2}$   \\
WD1323-514    & 13:26:09.65$^{1}$   & --51:41:35.78$^{1}$   & 14.60$^{2}$   & 14.60$^{2}$   & DA2$^{2}$   & WD2134+218    & 21:36:36.30$^{2}$   &  +22:04:33.00$^{2}$   & 14.41$^{2}$   & 14.45$^{2}$   & DA3$^{2}$   \\
WD1327-083    & 13:30:13.64$^{6}$   & --08:34:29.49$^{6}$   & 12.40$^{7}$   & 12.34$^{7}$   & DA4$^{2}$   & WD2140+207    & 21:42:42.00$^{1}$   &  +20:59:58.24$^{1}$   & 13.40$^{2}$   & 13.24$^{2}$   & DQ6$^{2}$   \\
GJ521         & 13:39:24.10$^{6}$   &  +46:11:11.37$^{6}$   & 11.50$^{9}$   & 10.26$^{9}$   & M2$^{12}$   & WD2147+280    & 21:49:54.53$^{1}$   &  +28:16:59.80$^{1}$   & 14.66$^{2}$   & 14.68$^{2}$   & DB4$^{2}$   \\
WD1408+323    & 14:10:26.95$^{1}$   &  +32:08:36.10$^{1}$   & 13.96$^{2}$   & 13.97$^{2}$   & DA3$^{2}$   & WD2152-548    & 21:56:21.27$^{1}$   & --54:38:23.00$^{1}$   & 13.80$^{2}$   & 14.30$^{2}$   & DA1$^{2}$   \\
SDSS~09626    & 14:29:51.06$^{3}$   &  +39:28:25.43$^{3}$   & 15.23$^{4}$   & 14.99$^{4}$   & A0$^{5}$    & GJ851         & 22:11:30.09$^{6}$   &  +18:25:34.29$^{6}$   & 11.37$^{7}$   & 10.23$^{7}$   & M2$^{12}$   \\
GJ570.2       & 14:57:32.30$^{6}$   &  +31:23:44.61$^{6}$   & 12.68$^{29}$  & 11.54$^{29}$  & M2$^{12}$   & WD2211-495    & 22:14:11.91$^{9}$   & --49:19:27.26$^{9}$   & 11.37$^{2}$   & 11.71$^{2}$   & DA1$^{2}$   \\
G15-10        & 15:09:46.02$^{6}$   & --04:45:06.61$^{6}$   & 12.67$^{26}$  & 12.01$^{26}$  & G2$^{41}$   & WD2216-657    & 22:19:48.35$^{1}$   & --65:29:18.11$^{1}$   & 14.57$^{2}$   & 14.43$^{2}$   & DZ5$^{2}$   \\
WD1509+322    & 15:11:27.66$^{1}$   &  +32:04:17.80$^{1}$   & 14.20$^{2}$   & 14.11$^{2}$   & DA3$^{2}$   & GJ863         & 22:33:02.23$^{6}$   &  +09:22:40.70$^{6}$   & 11.91$^{7}$   & 10.74$^{7}$   & M0$^{12}$   \\
M5-S1490      & 15:17:38.64$^{57}$  &  +02:02:25.60$^{57}$  & 15.08$^{57}$  & 14.10$^{57}$  & ---         & SDSS~14276    & 22:42:04.17$^{3}$   &  +13:20:28.61$^{3}$   & 14.48$^{4}$   & 14.32$^{4}$   & A0$^{5}$    \\ 
G167-50       & 15:35:31.55$^{1}$   &  +27:51:02.20$^{1}$   & 14.25$^{15}$  & 13.50$^{15}$  & G$^{42}$    & WD2251-634    & 22:55:10.00$^{2}$   & --63:10:27.00$^{2}$   & ---           & 14.28$^{2}$   & DA$^{2}$    \\
G179-54       & 15:46:08.25$^{1}$   &  +39:14:16.40$^{1}$   & 13.90$^{42}$  & 13.41$^{42}$  & F$^{42}$    & WD2309+105    & 23:12:21.62$^{25}$  &  +10:47:04.25$^{25}$  & 12.78$^{32}$  & 13.09$^{32}$  & DA1$^{2}$   \\ 
G224-83       & 15:46:14.68$^{1}$   &  +62:26:39.60$^{1}$   & 13.86$^{15}$  & 12.67$^{15}$  & K$^{42}$    & G190-15       & 23:13:38.82$^{6}$   &  +39:25:02.59$^{6}$   & 11.57$^{29}$  & 10.98$^{29}$  & F6$^{28}$   \\
WD1553+353    & 15:55:01.99$^{1}$   &  +35:13:28.70$^{1}$   & 14.64$^{2}$   & 14.75$^{2}$   & DA2$^{2}$   & SDSS~ 00832   & 23:30:24.90$^{3}$   & --00:09:34.90$^{3}$   & 15.15$^{4}$   & 14.99$^{4}$   & A0$^{5}$    \\ 
G16-20        & 15:58:18.62$^{9}$   &  +02:03:06.11$^{9}$   & 11.34$^{42}$  & 10.75$^{42}$  & K$^{20}$    & WD2329+407    & 23:31:35.65$^{1}$   &  +41:01:30.70$^{1}$   & 13.85$^{2}$   & 13.82$^{2}$   & DA3$^{2}$   \\   
WD1606+422    & 16:08:22.20$^{1}$   &  +42:05:43.20$^{1}$   & 13.93$^{2}$   & 13.82$^{2}$   & DA4$^{2}$   & WD2331-475    & 23:34:02.20$^{1}$   & --47:14:26.50$^{1}$   & 13.15$^{3}$   & 13.44$^{2}$   & DA1$^{2}$   \\
WD1615-154    & 16:17:55.26$^{1}$   & --15:35:51.90$^{1}$   & 13.22$^{2}$   & 13.42$^{2}$   & DA2$^{2}$   & G241-64       & 23:41:24.49$^{1}$   &  +59:24:34.90$^{1}$   & 13.45$^{15}$  & 12.70$^{15}$  & K$^{20}$    \\
GJ625         & 16:25:24.62$^{6}$   &  +54:18:14.77$^{6}$   & 11.80$^{9}$   & 10.17$^{9}$   & M2$^{12}$   & G171-15       & 23:45:02.71$^{9}$   &  +44:40:03.60$^{9}$   & 12.00$^{9}$   & 11.75$^{9}$   & G0$^{56}$   \\
G180-58       & 16:28:16.87$^{6}$   &  +44:40:38.28$^{6}$   & 11.87$^{29}$  & 11.12$^{29}$  & G/K$^{20}$  & WD2352+401    & 23:54:56.25$^{1}$   &  +40:27:30.10$^{1}$   & 15.13$^{2}$   & 14.94$^{2}$   & DQ6$^{2}$   \\
WD1626+368    & 16:28:25.03$^{1}$   &  +36:46:15.40$^{1}$   & 14.02$^{2}$   & 13.83$^{2}$   & DZA6$^{2}$  \\
WD1637+335    & 16:39:27.83$^{25}$  &  +33:25:22.30$^{25}$  & 14.85$^{2}$   & 14.65$^{2}$   & DA5$^{2}$   \\
SDSS~13028    & 16:40:24.18$^{3}$   &  +24:02:14.91$^{3}$   & 15.45$^{4}$   & 15.26$^{4}$   & A0$^{5}$    \\
WD1659-531    & 17:02:56.33$^{43}$  & --53:14:36.63$^{43}$  & 13.57$^{2}$   & 13.47$^{2}$   & DA4$^{2}$   \\
G139-16       & 17:09:47.38$^{1}$   &  +08:04:25.50$^{1}$   & 13.31$^{24}$  & 12.61$^{24}$  & K$^{20}$    \\
G170-47       & 17:32:41.63$^{6}$   &  +23:44:11.64$^{6}$   & 9.54$^{9}$    & 8.94$^{9}$    & G0$^{28}$   \\
2MASS~J175713 & 17:57:13.25$^{1}$   &  +67:03:40.90$^{1}$   & 11.91$^{1}$   & 12.01$^{1}$   & A3$^{44}$   \\
TYC4213-617   & 18:00:02.14$^{14}$  &  +66:45:54.96$^{14}$  & 11.24$^{9}$   & 10.68$^{9}$   & ---         \\   
BD+661071     & 18:02:10.92$^{14}$  &  +66:12:26.39$^{14}$  & 10.93$^{9}$   & 10.52$^{9}$   & F5$^{42}$   \\
G184-17       & 18:40:29.27$^{1}$   &  +19:36:06.65$^{1}$   & 14.90$^{27}$  & 14.08$^{27}$  & K$^{20}$    \\
WD1837-619    & 18:42:29.73$^{45}$  & --61:51:45.10$^{45}$  & 15.01$^{2}$   & 14.90$^{2}$   & DC5$^{2}$   \\
G184-20       & 18:43:52.50$^{1}$   &  +16:00:34.20$^{1}$   & 13.37$^{46}$  & 12.61$^{47}$  & G$^{20}$    \\
WD1845+019    & 18:47:39.08$^{1}$   &  +01:57:35.62$^{1}$   & 12.73$^{2}$   & 12.95$^{2}$   & DA2$^{2}$   \\
WD1900+705    & 19:00:10.25$^{1}$   &  +70:39:51.24$^{1}$   & 13.24$^{2}$   & 13.19$^{2}$   & DAP4$^{9}$  \\
GJ745A        & 19:07:05.56$^{6}$   &  +20:53:16.97$^{6}$   & 12.40$^{7}$   & 10.77$^{7}$   & M2$^{12}$   \\   
GJ745B        & 19:07:13.20$^{6}$   &  +20:52:37.24$^{6}$   & 12.38$^{7}$   & 10.77$^{7}$   & M2$^{12}$   \\   
WD1918+725    & 19:18:10.5$^{2}$    &  +72:37:24.00$^{2}$   & 14.70$^{48}$  & 15.12$^{48}$  & DA2$^{2}$   \\
WD1914-598    & 19:18:44.84$^{1}$   & --59:46:33.80$^{1}$   & 14.34$^{2}$   & 14.39$^{2}$   & DA$^{2}$    \\
WD1919+145    & 19:21:40.40$^{2}$   &  +14:40:43.00$^{2}$   & 13.07$^{2}$   & 13.01$^{2}$   & DA3$^{2}$   \\ 
WD1936+327    & 19:38:28.21$^{1}$   &  +32:53:19.90$^{1}$   & 13.46$^{2}$   & 13.58$^{2}$   & DA2$^{2}$   \\  
G23-14        & 19:51:49.61$^{6}$   &  +05:36:45.84$^{6}$   & 11.42$^{49}$  & 11.02$^{49}$  & G5$^{26}$   \\    
WD2000-561.1  & 20:04:18.00$^{2}$   & --56:02:47.00$^{2}$   & ---           & 15.20$^{50}$  & DA1$^{2}$   \\  
WD2004-605    & 20:09:05.24$^{51}$  & --60:25:41.60$^{51}$  & 13.10$^{2}$   & 13.40$^{2}$   & DA1$^{2}$   \\  
WD2014-575    & 20:18:54.90$^{52}$  & --57:21:34.00$^{52}$  & 13.40$^{2}$   & 13.70$^{2}$   & DA2$^{2}$   \\
WD2028+390    & 20:29:56.16$^{1}$   &  +39:13:32.00$^{1}$   & 13.22$^{2}$   & 13.37$^{2}$   & DA2$^{2}$   \\  
WD2032+248    & 20:34:21.88$^{6}$   &  +25:03:49.72$^{6}$   & 11.47$^{7}$   & 11.55$^{7}$   & DA2$^{2}$   \\  
WD2034-532    & 20:38:16.84$^{1}$   & --53:04:25.40$^{1}$   & 14.41$^{2}$   & 14.46$^{2}$   & DB4$^{2}$   \\  
G24-25        & 20:40:16.10$^{9}$   &  +00:33:19.74$^{9}$   & 11.23$^{9}$   & 10.61$^{9}$   & G0$^{53}$   \\ 
WD2039-202    & 20:42:34.75$^{6}$   & --20:04:35.95$^{6}$   & 12.32$^{7}$   & 12.40$^{7}$   & DA3$^{2}$   \\ 
SDSS~14511    & 20:42:42.40$^{3}$   & --00:34:03.71$^{3}$   & 15.34$^{4}$   & 15.11$^{4}$   & A0/F0$^{5}$ \\
WD2039-682    & 20:44:21.47$^{54}$  & --68:05:21.30$^{54}$  & 13.19$^{2}$   & 13.25$^{2}$   & DA3$^{2}$   \\ 
SDSS~15724    & 20:47:38.19$^{3}$   & --06:32:13.11$^{3}$   & 15.06$^{4}$   & 14.87$^{4}$   & A0/F2$^{5}$ \\ 
\end{longtable}


\label{lastpage}

\end{document}